\def\Cu{$\kappa$-(BEDT\--TTF)$_2$\-Cu$_2$(CN)$_3$}

\def\Cl{$\kappa$-(BEDT\--TTF)$_2$\-Cu[N(CN)$_2$]Cl}

\def\Pdmit{EtMe$_3$\-P[Pd(dmit)$_2$]$_2$}

\documentclass[aps,prb,twocolumn,showpacs,superscriptaddress,am,citeautoscript]{revtex4-2}
\usepackage{graphicx}
\usepackage{amsmath}
\usepackage{amssymb}
\usepackage{color}

\usepackage{hyperref}

\setcitestyle{super}

\begin{document} 
\title{Chasing the spin gap through the phase diagram of a frustrated Mott insulator}

\author{A. Pustogow}
\affiliation{Institute of Solid State Physics, TU Wien, 1040 Vienna, Austria}
\author{Y. Kawasugi}
\affiliation{Department of Physics, Toho University, Funabashi 274-8510, Chiba, Japan}
\affiliation{Condensed Molecular Materials Laboratory, RIKEN, Wako, Saitama 351-0198, Japan}
\author{H. Sakurakoji}
\affiliation{Department of Physics, Toho University, Funabashi 274-8510, Chiba, Japan}
\author{N. Tajima}
\affiliation{Department of Physics, Toho University, Funabashi 274-8510, Chiba, Japan}
\affiliation{Condensed Molecular Materials Laboratory, RIKEN, Wako, Saitama 351-0198, Japan}
%
%
%
%
\begin{abstract}

The quest for entangled spin excitations 
has stimulated intense research on frustrated magnetic systems. 
For almost two decades, the triangular-lattice Mott insulator \Cu\ has been the hottest candidate for a $gapless$ quantum spin liquid with itinerant spinons. Very recently, however, this scenario was overturned as electron-spin-resonance (ESR) studies unveiled a spin gap, calling for reevaluation of the magnetic ground state. 
Here we achieve a precise mapping of this spin-gapped phase through the Mott transition by ultrahigh-resolution strain tuning. Our transport experiments reveal a reentrance of charge localization below $T^{\star}=6$~K associated with a gap size of 30--50~K. The negative slope of the insulator-metal boundary, $dT^{\star}/dp<0$, evidences the low-entropy nature of the spin-singlet ground state. 
By tuning the enigmatic '6K anomaly' through the phase diagram of \Cu, we identify it as the transition to a valence-bond-solid phase, with typical magnetic and structural fingerprints, that persists at $T\rightarrow 0$ until unconventional superconductivity and metallic transport proliferate.

\end{abstract}
%


\maketitle
%
%

Since Anderson's notion of resonating valence bonds~\cite{Anderson1973}, the quest for quantum spin liquids (QSL) has been fueled by the idea that suppressing long-range antiferromagnetic (AFM) order can stabilize itinerant or even topological spin excitations~\cite{Balents2010,Broholm2020}. What is often neglected  
is that the coupling of magnetic degrees of freedom to the lattice can result in valence-bond-solid (VBS) phases with paired electron spins. Such singlet states occur, for instance, in triangular-lattice organic compounds~\cite{Tamura2006,Shimizu2007,Itou2009,Manna2014,Yoshida2015},
and in form of the well-known spin-Peierls (SP) states  
in quasi 1D systems~\cite{Dumm2000,Hase1993}.  
Also in other frustrated materials the importance of valence-bond phases has been revived~\cite{Hermanns2015,Norman2020,Kimchi2018,Riedl2019}. 
Such a VBS scenario is fully consistent with the report of a spin gap in the organic QSL candidate \Cu\ that aroused great attention in the QSL community very recently \cite{Miksch2021,Pustogow2022}. While this finding rules out the widely assumed scenario of mobile spinons~\cite{Yamashita2008}, in agreement with thermal transport results~\cite{Yamashita2009}, it leaves open the possibility of gapped QSL phases~\cite{Broholm2020}.

\begin{figure*}
 \centering
 \includegraphics[width=2.05\columnwidth]{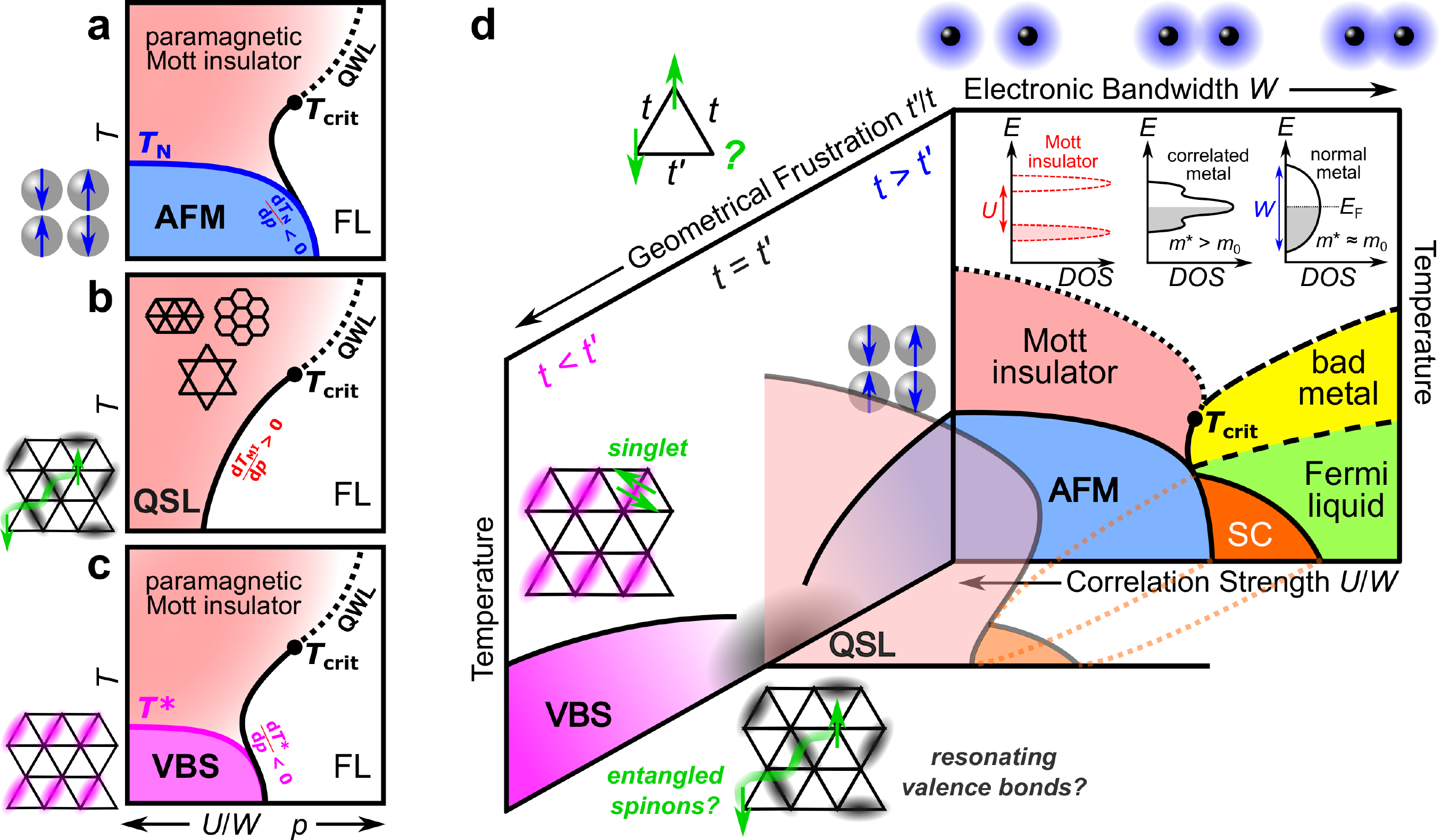}
 \caption{\footnotesize \textbf{a} While the entropy of a paramagnetic Mott insulator exceeds that of the adjacent Fermi liquid (FL), resulting in a positive slope $dT_{\rm MI}/dp>0$ of the metal-insulator boundary, AFM order has much smaller entropy and the Clausius-Clapeyron relation yields $dT_{\rm N}/dp<0$, as seen in \Cl\  \cite{Lefebvre2000,Limelette2003a}. \textbf{b} No such 'reentrance' of insulating behavior is expected for a $gapless$ QSL, possibly realized in triangular, kagome or honeycomb lattices~\cite{Balents2010,Broholm2020}.
 \textbf{c} Similar to AFM, also the transition from a spin-gapped VBS insulator to a metal yields $dT^{\star}/dp<0$, for instance in \Pdmit\ \cite{Shimizu2007,Itou2009}. 
 \textbf{d} Geometrical frustration $t'/t$ allows to tune the magnetic ground state of $genuine$ Mott insulators, causing pronounced changes in the phase diagram affecting also unconventional superconductivity (SC).  
 }
\label{phase-diagram}
\end{figure*}

The tunability of \Cu\ -- a pressure of $1.3$~kbar triggers the transition from a Mott insulator to superconducting (SC) and metallic phases~\cite{Kurosaki2005,Shimizu2007,Itou2009,Shimizu2016,Furukawa2018,Pustogow2018} as illustrated in Fig.~\ref{phase-diagram}d -- provides the unique opportunity to scrutinize the entropy of its magnetic ground state. According to the Clausius-Clapeyron relation, the pressure-tuned metal-insulator transition (MIT) acquires a positive slope $dT_{\rm MI}/dp>0$ in the $T$-$p$ phase diagram between a paramagnetic Mott insulator with large spin entropy and a metal (Fig.~\ref{phase-diagram}b), reminiscent of the Pomeranchuk effect in $^3$He~\cite{Pustogow2018}. 
On the other hand, the entropy of AFM and VBS states is smaller than that of the Fermi-liquid (FL) metal, yielding negative slopes $dT_{\rm N}/dp<0$ and $dT^{\star}/dp<0$ of the MIT (Fig.~\ref{phase-diagram}a,c), respectively.

Here we utilize the slope of the insulator-metal boundary of \Cu\ as a probe of the spin-gapped ground state below $T^{\star}=6$~K, evidencing its low-entropy nature. 
For that, we map the low-temperature Mott MIT with ultrahigh precision by measuring dc transport upon a combination of biaxial compression and tensile strain. We find a reentrance of insulating behavior setting in below $T^{\star}$, which is successively suppressed as metallicity proliferates.
Through comparison with previously reported structural~\cite{Manna2010,Manna2014} and magnetic~\cite{Miksch2021} properties, our present results clearly identify the '6K anomaly' as the transition to a VBS phase with $dT^{\star}/dp<0$ (Fig.~\ref{phase-diagram}c) and, thus, we complete the phase diagram of \Cu. Moreover, our transport data yield a gap size of 30--50~K for the spin-singlet state, which coincides with the unexpectedly large critical field of 60~T \cite{Pustogow2022}. 

Among the QSL candidates, \Cu\ has been studied most intensely as it exhibits not only frustrated AFM exchange interactions ($J=250$~K~\cite{Shimizu2003}), but also a paradigmatic Mott transition~\cite{Kurosaki2005,Furukawa2015,Pustogow2018}. This layered triangular-lattice compound with $t'/t$ close to unity 
shows no AFM order~\cite{Shimizu2003} and, at the same time, it is one of the best solid-state realizations of the single-band Hubbard model~\cite{Pustogow2018}. As such, it provides experimental access to check the predictions of dynamical mean-field theory: while the Mott transition is first-order type below the critical endpoint $T_{crit}$, a quantum-critical crossover at the quantum Widom line (QWL) occurs at $T>T_{crit}$~\cite{Terletska2011,Vucicevic2013,Furukawa2015,Pustogow2018}. Moreover, applying 1--2~kbar pressure allows to probe unconventional SC~\cite{Kurosaki2005,Furukawa2018} as well as charge transport in Fermi liquids and $bad$ metals~\cite{Pustogow2021-Landau}. By combining the most recent findings in \Cu\ \cite{Furukawa2015,Pustogow2018,Pustogow2021-Landau} and compounds with different degree of frustration~\cite{Lefebvre2000,Limelette2003a,Shimizu2007,Furukawa2015,Shimizu2016,Yoshida2015}, in Fig.~\ref{phase-diagram}d we present the state-of-the-art phase diagram of Mott systems as a function of $T$, electronic correlations $U/W$ and geometrical frustration $t'/t$.

\begin{figure}
 \centering
 \includegraphics[width=0.98\columnwidth]{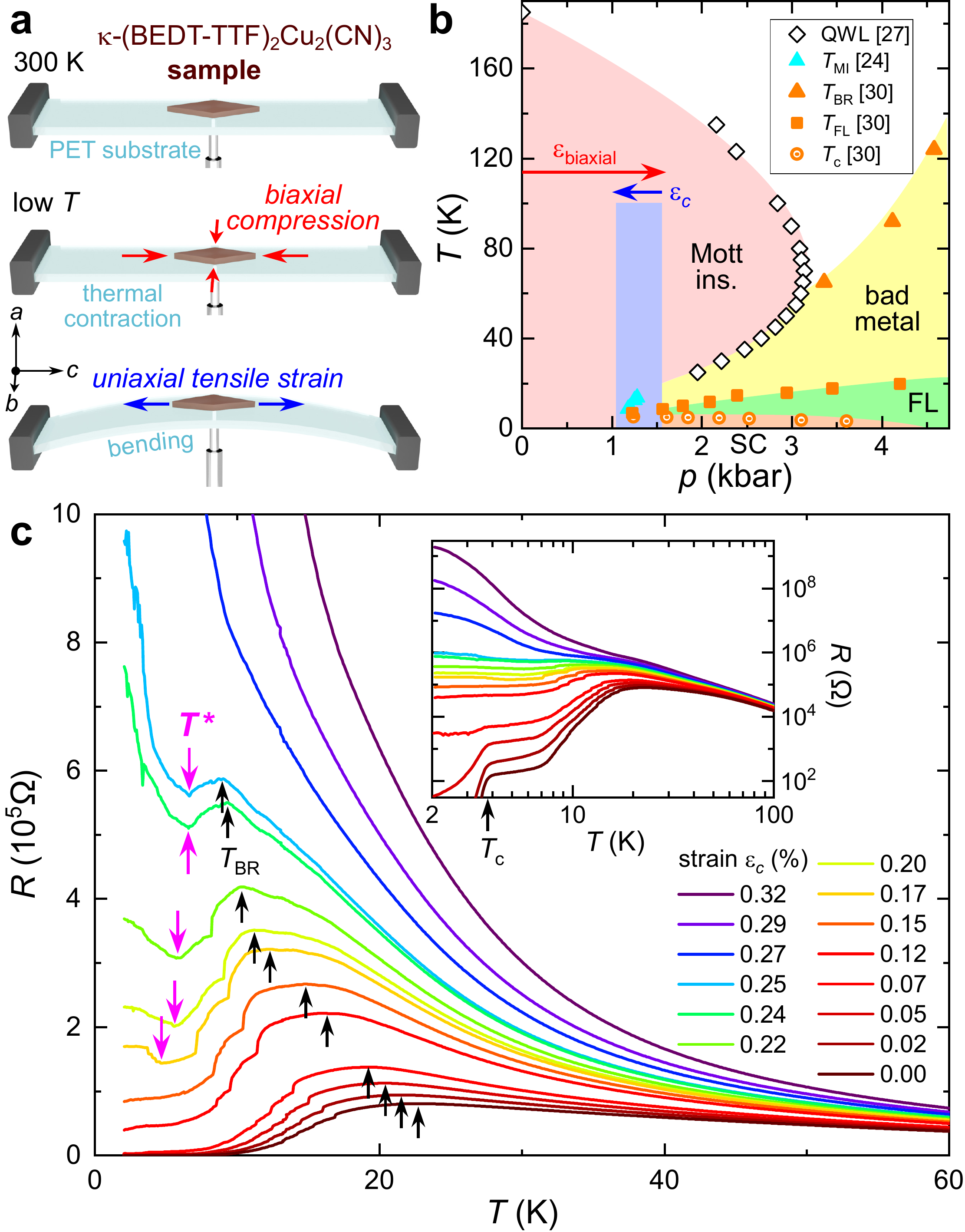}
 \caption{\footnotesize \textbf{a,b} A 140 nm thin single-crystalline filament of \Cu\ was placed on a flexible PET substrate. Cooling from 300~K down to $T\rightarrow 0$ imposes biaxial compression of order 1--2~kbar to the sample due to differential thermal contraction, yielding metallic and superconducting (SC) properties (red arrow in \textbf{b}). Uniaxial tensile strain $\varepsilon_c$ along the $c$-axis is applied via bending of the substrate in order to tune from the Fermi liquid (FL) through the Mott MIT (blue arrow in \textbf{b}). \textbf{b} Crossover and transition temperatures from Refs.~\cite{Furukawa2018,Furukawa2015,Pustogow2021-Landau}, as indicated. \textbf{c} Resistance of the biaxially compressed crystal for different $\varepsilon_c$. Inset: SC below $T_c\approx 4$~K for $\varepsilon_c=0$ is evident in the log-log plot. Upon approaching the Mott state, SC is lost and the resistivity maxima at $T_{\rm BR}$ shift to lower temperatures. In addition, our data reveal a reentrance of insulating behavior (magenta arrows) below $T^{\star}<T_{\rm BR}$. 
 }
\label{resistance}
\end{figure}

In this study, we explore the phase space extremely close to the MIT, which requires fine tuning of the correlation strength with a resolution equivalent to $10^{-2}$~kbar. While such high precision cannot be reached in conventional oil pressure cells~\cite{Kurosaki2005,Pustogow2021-Landau}, also gas pressure experiments are limited at low temperatures due to solidification of helium~\cite{Furukawa2015,Furukawa2018}. To that end, we utilize strain transmitted through a substrate -- a method previously applied in doping-tuned experiments~\cite{Suda2014}.
As sketched in Fig.~\ref{resistance}a, a single-crystalline \Cu\ filament of 140 nm thickness is placed on a flexible polyethylene substrate at ambient conditions. Differential thermal expansion between sample and substrate (see Methods) causes an in-plane biaxial 
compression $\varepsilon_{\rm biaxial}$ equivalent to 1--2~kbar hydrostatic pressure, which is sufficient to push the sample across the Mott MIT (Fig.~\ref{resistance}b). Our dc transport results in Fig.~\ref{resistance}c yield metallic properties and an onset of SC below $T_c\approx 4$~K ($\varepsilon_c=0$, see inset). Note, uniaxial strain experiments on \Cu\ revealed a considerable enhancement of $T_c$ and a broadened SC dome ~\cite{Shimizu2011} than for isotropic compression. Here, biaxial strain does not affect the in-plane anisotropy and $T_c$ is comparable to hydrostatic pressure experiments~\cite{Kurosaki2005,Furukawa2018}. 

\begin{figure}
 \centering
 \includegraphics[width=0.95\columnwidth]{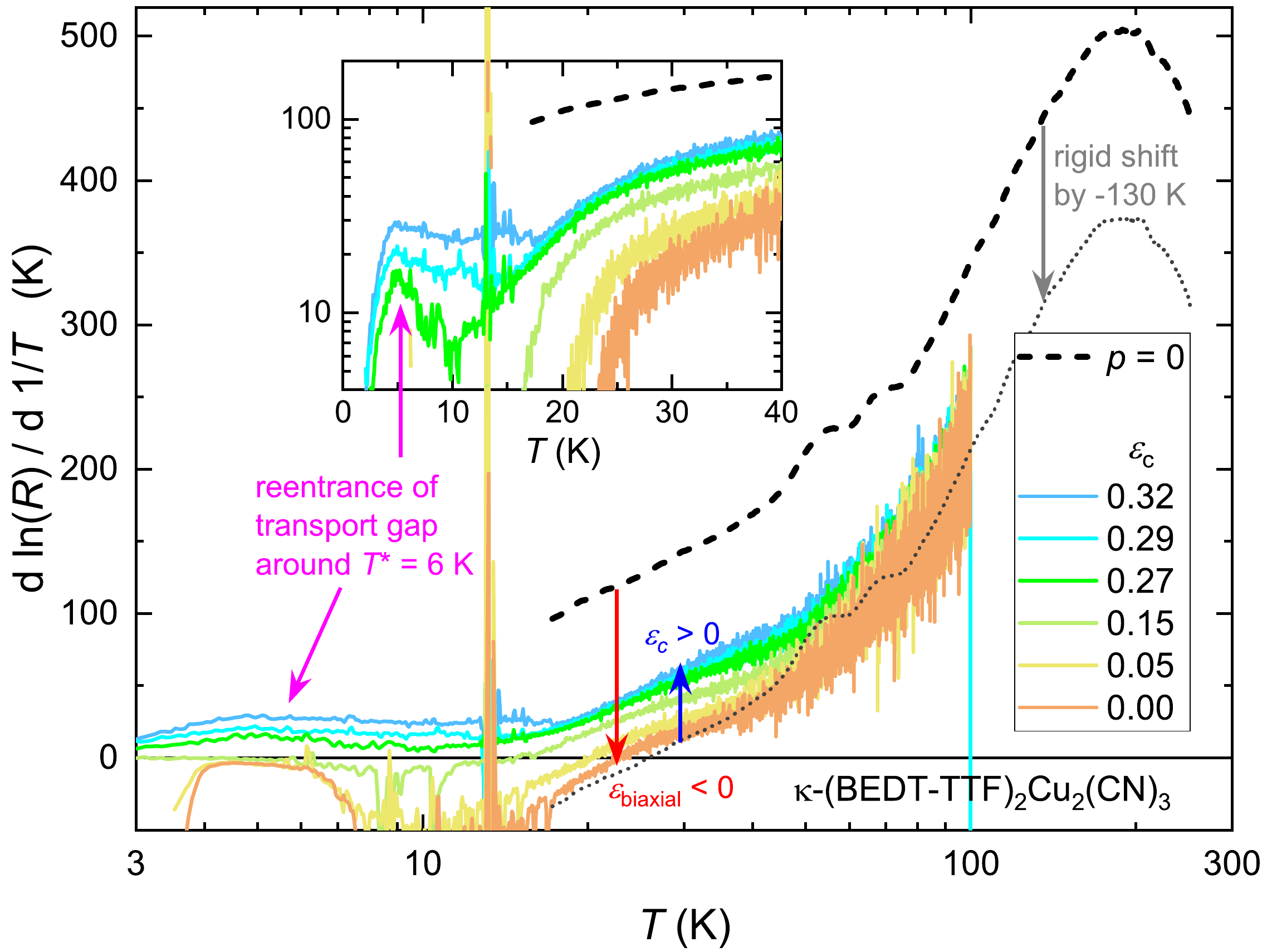}
 \caption{\footnotesize The transport gap $d\ln{R}/d\; 1/T$ determined from the strain-dependent resistance (Fig.~\ref{resistance}c) is compared to ambient pressure (black dashed line from 
 Ref.~\cite{Pustogow2018}). While biaxial compression $\varepsilon_{\rm biaxial}<0$ causes a reduction of the gap size, equivalent to a vertical shift of the $p=0$ data by -130~K (grey dotted), the gap increases upon tensile strain $\varepsilon_c>0$, consistent with the bandwidth-tuning in Fig.~\ref{resistance}b. The Mott MIT is evident from the sign change at $T_{\rm BR}$. Upon further cooling in the metallic state, $d\ln{R}/d\; 1/T \rightarrow 0$ as the residual resistivity is approached, followed by a sharp drop when SC sets in. Notably, the fully insulating curves exhibit a local maximum around $T^{\star}=6$~K associated with the VBS transition.
 }
\label{transport-gap}
\end{figure}

While differential thermal contraction  
tunes the sample across the Mott MIT, it does not allow to vary the correlation strength in arbitrary steps. In order to reach back into the Mott state in a quasi-continuous manner, we perform bandwidth-tuning via tensile strain $\varepsilon_c>0$ along the crystallographic $c$-axis. As indicated on the bottom of Fig.~\ref{resistance}a, we control the correlation strength by varying the bending radius of the substrate, which yields a $negative$ pressure of less than a kilobar.
As such, uniaxial strain is significantly weaker than the overall pressure applied ($\lvert\varepsilon_c\rvert<\lvert\varepsilon_{\rm biaxial}+\varepsilon_c\rvert$), hence we do not expect major effects from a change in frustration $t'/t$
\cite{Shimizu2011}.   
The blue shaded region in Fig.~\ref{resistance}b indicates the phase space covered in our experiments with a fixed $\varepsilon_{\rm biaxial}<0$ and varying tensile strain between $\varepsilon_c=0$ and $0.32\%$. The resulting transport curves in Fig.~\ref{resistance}c exhibit a textbook Mott MIT: the resistivity maxima at the Brinkman-Rice temperature $T_{\rm BR}$ are pushed to lower temperatures as correlations increase~\cite{Pustogow2021-Landau,Kurosaki2005}. 
In addition, in the range between  $\varepsilon_c = 0.17\%$ and 0.25\% we observe a distinct upturn of resistance upon cooling well below $T_{\rm BR}$. The resistance minima (magenta arrows in Fig.~\ref{resistance}c) shift up to $T^{\star}=6$~K until metallic behavior is lost completely. 
Since this phenomenology lines up with the '6K anomaly' at ambient pressure, we identify it as the transition to a spin-gapped nonmetallic state that gets suppressed at the Mott MIT.

The reentrant behavior at low temperatures is apparent not only from a sign change of $dR/dT$ to a non-metallic slope ($dR/dT<0$ below $T^{\star}$), but can be seen also in the fully insulating curves $\varepsilon_c \geq 0.27\%$. In Fig.~\ref{transport-gap} we plot the logarithmic derivative $d\ln{R}/d\; 1/T$ as a measure of the transport gap $\Delta (T)$. While the gap size is steadily reduced upon cooling from 100~K to 20~K, consistent with the temperature-dependent spectral weight shifts reported in previous optical studies~\cite{Pustogow2018,Pustogow2021-Landau}, 
we observe an upturn  at low temperatures forming a peak around $T^{\star}$. 
For comparison, 
the VBS transition of \Pdmit\ at $T^{\star}=24$~K yields similar maxima in $d\ln{R}/d\; 1/T$ from which we estimate a ratio of gap size and transition temperature $\Delta_{VBS}/(k_B T^{\star})\approx 5$~\cite{Shimizu2007}. Assuming the same ratio in the title compound gives $\Delta_{VBS}/k_B\approx 30$~K for $T^{\star}=6$~K, which compares well with the local maxima in the inset of Fig.~\ref{transport-gap}.
In ambient-pressure transport studies ~\cite{Pustogow2018}, the regime $T<10$~K remained inaccessible due to the extremely large resistance; here, in the vicinity of the MIT $R(T)$ is small enough to be measured. Following its observation in thermodynamic and magnetic probes~\cite{Yamashita2008,Manna2010,Miksch2021}, in our present work we identify the '6K anomaly' in the Mott-insulating state of \Cu\ for the first time in charge transport as a local maximum of $d\ln{R}/d\; 1/T$. Even more importantly, we obtain a quantitative estimate of the associated energy gap $E_g=2\Delta\approx 30$--50~K at $T^{\star}$. Note, this significantly exceeds the spin gap of 12~K estimated from ESR data
~\cite{Miksch2021}. 
Crucially, the gap size observed here lines up with the critical field of order 60~T estimated in Ref.~\cite{Pustogow2022} , which explains why magnetic fields of order $10$~T 
 have little effect on the transition~\cite{Yamashita2008,Manna2010}.

\begin{figure*}
 \centering
 \includegraphics[width=2.05\columnwidth]{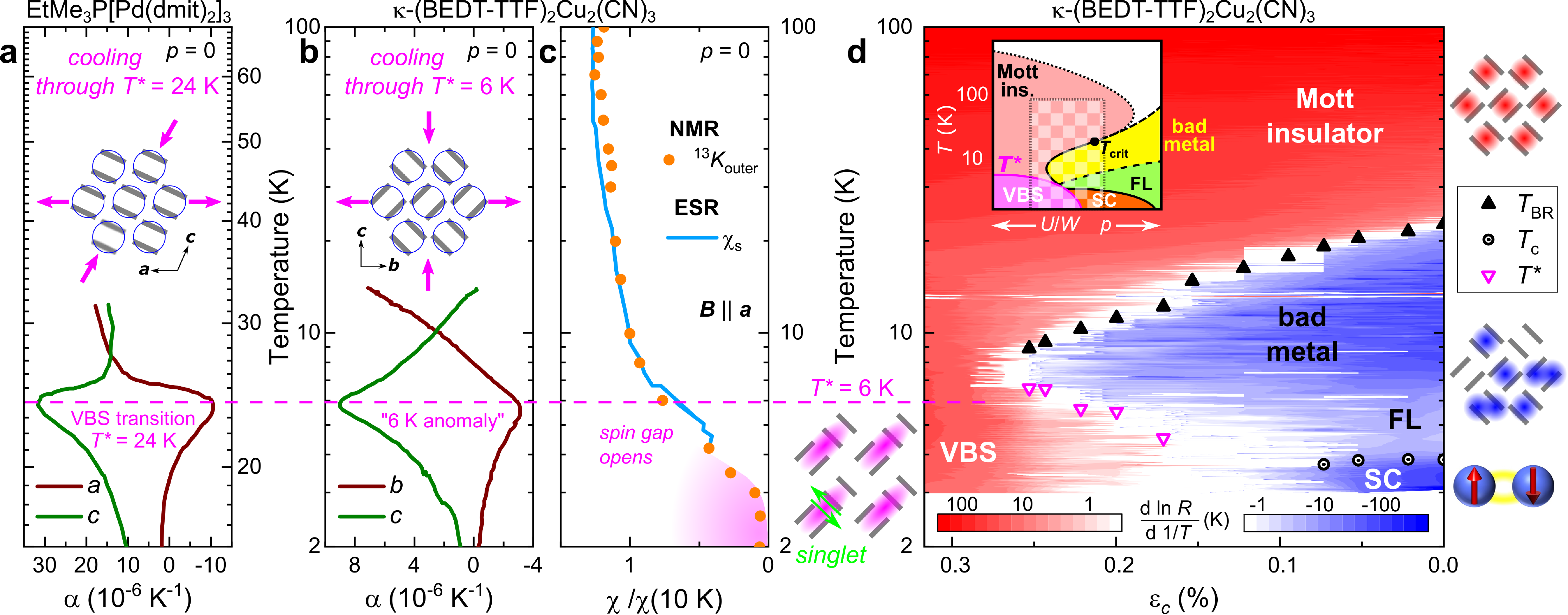}
 \caption{\footnotesize \textbf{a,b} The thermal expansion coefficient $\alpha$ yields similar in-plane anisotropy at the VBS transition of \Pdmit\  ($T^{\star}=24$~K) \cite{Manna2014} and at the '6K anomaly' of \Cu\ \cite{Manna2010}. Note the different scales; data from Refs.~\cite{Manna2010,Manna2014} .
 \textbf{c} The rapid drop of NMR Knight shift $K$ and ESR susceptibility $\chi_s$ ($B\parallel a$) at $T^{\star}=6$~K evidences the opening of a spin gap; data from Refs.~\cite{Miksch2021,Saito2018} . 
 \textbf{d} The phase diagram of \Cu\ assembled from $R(T,\varepsilon_c)$ (Figs.~\ref{resistance}c,~\ref{transport-gap}) is underlaid with a false-color plot of $d\ln{R}/d\; 1/T$ (red: insulating, blue: metallic). Inset: checkerboard area indicates range of our strain-dependent experiments. We find a clear back-bending of the insulator-metal boundary (cf. Fig.~\ref{phase-diagram}c) originating from the '6K anomaly' at $p=0$. 
 Altogether, the ambient pressure data in \textbf{b,c} and $dT^{\star}/dp<0$ in \textbf{d} provide solid evidence for a low-entropy spin-singlet phase with structural anisotropy -- the hallmarks of a VBS. 
 }
\label{6K-anomaly}
\end{figure*}

$d\ln{R}/d\; 1/T$ in the range $T> 20$~K reflects changes of the Mott-Hubbard gap.  
While the gap size reduces upon biaxial compression, it increases upon uniaxial tension; the negative values below $T_{\rm BR}$ correspond to $dR/dT>0$.  
Interestingly, the strain-induced change upon $\varepsilon_{\rm biaxial}<0$ is approximately a rigid vertical shift of the ambient-pressure data~\cite{Pustogow2018} 
by $-130$~K (grey dotted line in Fig.~\ref{transport-gap}). Quantitatively, the effect of $\varepsilon_{\rm biaxial}$ is approximately three times bigger than $\varepsilon_c =0.32\%$, consistent with the red and blue arrows in Fig.~\ref{resistance}b, respectively.

In Fig.~\ref{6K-anomaly}a,b we compare the structural changes at the '6K anomaly' in \Cu\ with the VBS transition at $T^{\star}=24$~K in \Pdmit\ \cite{Tamura2006}. Thermal expansion experiments by Manna \textit{et al.}~\cite{Manna2010,Manna2014} revealed very similar anisotropic distortions within the conducting layers at $T^{\star}$ in both quasi 2D organic compounds. As sketched in Fig.~\ref{6K-anomaly}a,b, the crystal contracts along the $c$-axis and expands along the other in-plane direction upon cooling. Also 
other probes~\cite{Itoh2013,Kobayashi2020} observed pronounced anomalies at 6~K due to modifications of the crystal structure and possible symmetry breaking~\cite{Pustogow2022}, which are typical fingerprints of a VBS transition. 
In addition, recent ESR studies revealed a rapid drop of spin susceptibility $\chi_s$ due to the opening of a spin gap~\cite{Miksch2021}, which agrees with NMR Knight shift data from Ref.~\cite{Saito2018} that yield $\chi_s$ indistinguishable from zero for $T\rightarrow 0$, as shown in Fig.~\ref{6K-anomaly}c.

Altogether, these findings provide solid evidence for anisotropic structural changes at the transition to a singlet state with a gap in the spectrum of spin excitations. In addition, $T^{\star}=6$~K at $p=0$ directly interpolates to the reentrance of the MIT in our strain-dependent transport data, which is mapped in the false-color plot of $d\ln{R}/d\; 1/T$ in Fig.~\ref{6K-anomaly}d. We find that the resulting phase diagram prominently displays $dT^{\star}/dp<0$ as sketched in Fig.~\ref{phase-diagram}c, analogue to the VBS Mott insulator \Pdmit\ \cite{Shimizu2007,Itou2009} with similar thermal expansion anomalies (Fig.~\ref{6K-anomaly}a,b). 
In light of these overwhelming experimental similarities, we conclude that at ambient pressure the enigmatic '6K anomaly' in \Cu\  is the transition from a paramagnetic Mott insulator to a spin-gapped VBS state. Upon reducing the correlation strength, the VBS phase is suppressed and can be entered from a metallic state upon cooling, as observed here for $0.17\% \leq \varepsilon_c \leq 0.25\% $. As seen in Fig.~\ref{6K-anomaly}d, the nonmagnetic insulator is adjacent to unconventional SC in the limit $T\rightarrow 0$, suggesting that the '6K anomaly' is relevant for Cooper pair formation.

Since the first reports of the '6K anomaly' and the absence of magnetic order~\cite{Shimizu2003,Yamashita2008}, there have been different interpretations of the experimental results acquired 
 over two decades in independent groups~\cite{Yamashita2009,Manna2010,Saito2018,Miksch2021}. 
 Here, we put together now all these pieces and obtain a comprehensive understanding of the low-temperature physics of \Cu. Our present ultrahigh-resolution tuning of the '6K anomaly' through the Mott MIT delivers the missing part required to establish it as a true phase transition to a low-entropy spin-singlet ground state with distorted crystal structure. In view of the very shape of the phase diagram presented here, the magnetic, thermal transport, thermodynamic and structural properties of \Cu\ \cite{Shimizu2003,Yamashita2009,Manna2010,Saito2018,Miksch2021} 
are consistent with a spin-gapped VBS state, essentially identical to that reported in other layered organic Mott systems~\cite{Tamura2006,Shimizu2007,Itou2009,Manna2014,Yoshida2015}.

As such, we point out that the spin-gapped low-entropy phase associated with the '6K anomaly' is not intrinsic to $pure$ Mott insulators, but rather results from magneto-structural instabilities. Likely, the involved elastic energy 
contributes to the unexpectedly large gap and critical field, exceeding $k_B T^{\star}/\mu_B \approx 9$~T by far~\cite{Pustogow2022}. 
We suggest that, in absence of the VBS, the metallic state persists much deeper into the insulating region of the phase diagram -- in Fig.~\ref{resistance}b $T_{\rm FL}$ and $T_{\rm BR}$ extrapolate to zero well below $p=1$~kbar~\cite{Pustogow2021-Landau,Furukawa2018}. To that end, elucidating the $genuine$ Mott MIT in the limit $T\rightarrow 0$ requires to suppress $T^{\star}$ in \Cu, potentially by varying $t'/t$ through uniaxial strain (Fig.~\ref{phase-diagram}d) or by applying high magnetic fields $B\approx 50$--100~T.
To conclude, our findings highlight the importance to gain understanding and control of magneto-elastic coupling in frustrated spin systems, in particular in other QSL candidates.

\section*{Methods}

\subsection*{Sample preparation and application of uniaxial bending strain}
We prepared polyethylene terephthalate (PET) substrates (Teflex\textsuperscript{\textregistered\ }  FT7, Toyobo Film Solutions Limited) and patterned 18-nm-thick Au electrodes using photo-lithography.
Thin single crystals of $\kappa$-(BEDT-TTF)$_{2}$Cu$_{2}$(CN)$_{3}$ were synthesized with electrolysis of a 1,1,2-trichloroethane [10\% (v/v) ethanol] solution (50 ml), in which BEDT-TTF (20 mg), KCN (60mg), CuCN (40 mg), and 18-crown-6 (100 mg) are dissolved. We applied a current of 8 $\mu$A overnight and obtained tiny thin crystals of $\kappa$-(BEDT-TTF)$_{2}$Cu$_{2}$(CN)$_{3}$. The crystals were pipetted to 2-propanol, and were manipulated with a tip of hair and put on the substrate. After the substrate was taken out from 2-propanol and dried, the thin crystal tightly adhered to the substrate.

As described in previous work~\cite{Suda2014,Kawasugi2019,Kawasugi2021}, cooling yields differential thermal expansion between sample and substrate: the length change is $\leq 1\%$ between 0~K and 300~K for \Cu\ ~\cite{Manna2010,Jeschke2012} compared to $\gg 1\%$ for PET. As a result, the sample is subject to biaxial compression of order 1--2~kbar, deduced from the transport results in Fig.~\ref{resistance} in comparison to previously published resistivity data~\cite{Furukawa2015,Furukawa2018,Pustogow2021-Landau}.
We applied uniaxial bending strains by pushing the back of the substrate with a nanopositioner (ANPz51, attocube systems) as shown in Fig. 2a. Assuming that the bent substrate is an arc of a circle, the strain $S$ is estimated as $S=4tx/(l^{2}+4x^{2})$ \cite{Kawasugi2019,Kawasugi2021} using the small angle approximation, where $x$ is the displacement of the nanopositioner, $t=177 {\rm \mu m}$ and $l=12 {\rm mm}$ are the thickness and length of the substrate, respectively. The uniaxial tensile strain was applied at 100 K in descending order from 0\% to 0.32\%. The cooling rate of the sample was 0.5 K/min. The thickness of the studied \Cu\ crystal (140~nm) was measured \textit{a posteriori} using a step profiler. 
Throughout the manuscript we follow the convention that a reduction of the sample volume (upon compressive stress or pressure) corresponds to negative strain $\varepsilon < 0$, whereas an expansion of the sample size (upon tensile stress or 'negative' pressure) yields $\varepsilon > 0$.

\subsection*{Data Analysis}

The QWL in Fig.~\ref{resistance}b was taken from Ref.~\cite{Pustogow2018} whereas $T_{\rm BR}$, $T_{c}$ and $T_{\rm FL}$ are from Ref.~\cite{Pustogow2021-Landau} and the insulator-metal boundary below 20~K (cyan) was determined from the resistivity data in Ref.~\cite{Furukawa2018} .
The transport gap of \Cu\ at ambient pressure in Fig.~\ref{transport-gap} (dashed black line) was taken from Fig.~S2 in the Supplementary Information of Ref.~\cite{Pustogow2018} ; the grey dotted line is a rigid vertical shift of the ambient-pressure data by $-130$~K:  $\Delta(\varepsilon_{\rm biaxial},T)\approx\Delta(p=0,T)-130$~K.
Thermal expansion data in Fig.~\ref{6K-anomaly}a are from Ref.~\cite{Manna2014} and the data in panel b are from Ref.~\cite{Manna2010} .
The NMR shifts in Fig.~\ref{6K-anomaly}c were taken from Ref.~\cite{Saito2018} and the Knight shift  $K_{outer}$ was determined for the outer $^{13}$C nuclei using a chemical shift $\sigma=117$, as described in Ref.~\cite{Pustogow2022} . In the same panel, the ESR susceptibility $\chi_s$ for magnetic field perpendicular to the layers ($B\parallel a$ similar to the Knight shift data) was taken from Fig.~2a in Ref.~\cite{Miksch2021} . Both $\chi_s$ and $K_{outer}$ were normalized to their value at 10 K. 

At elevated temperatures the resistivity maxima indicate $T_{\rm BR}$~\cite{Pustogow2021-Landau}. At lower temperatures in the phase coexistence regime $T\leq T_{crit}\approx 15$~K~\cite{Pustogow2021-percolation}, strictly speaking, the resistivity maxima indicate percolation of metallic regions.
We assign the sharp jumps in the resistance data in Fig.~\ref{resistance}c, also seen as narrow peaks between 10--15~K in Fig.~\ref{transport-gap}, to an artefact of the sample mounting. As such, the apparent strain dependence is naturally related to the volume change of the sample that interferes with the applied strain. A micro-crack will open or close for a particular differential thermal expansion between sample and substrate, reached at a different temperature for different strain conditions upon changing $\varepsilon_c$.


\begin{thebibliography}{39}%
\makeatletter
\providecommand \@ifxundefined [1]{%
 \@ifx{#1\undefined}
}%
\providecommand \@ifnum [1]{%
 \ifnum #1\expandafter \@firstoftwo
 \else \expandafter \@secondoftwo
 \fi
}%
\providecommand \@ifx [1]{%
 \ifx #1\expandafter \@firstoftwo
 \else \expandafter \@secondoftwo
 \fi
}%
\providecommand \natexlab [1]{#1}%
\providecommand \enquote  [1]{``#1''}%
\providecommand \bibnamefont  [1]{#1}%
\providecommand \bibfnamefont [1]{#1}%
\providecommand \citenamefont [1]{#1}%
\providecommand \href@noop [0]{\@secondoftwo}%
\providecommand \href [0]{\begingroup \@sanitize@url \@href}%
\providecommand \@href[1]{\@@startlink{#1}\@@href}%
\providecommand \@@href[1]{\endgroup#1\@@endlink}%
\providecommand \@sanitize@url [0]{\catcode `\\12\catcode `\$12\catcode
  `\&12\catcode `\#12\catcode `\^12\catcode `\_12\catcode `\%12\relax}%
\providecommand \@@startlink[1]{}%
\providecommand \@@endlink[0]{}%
\providecommand \url  [0]{\begingroup\@sanitize@url \@url }%
\providecommand \@url [1]{\endgroup\@href {#1}{\urlprefix }}%
\providecommand \urlprefix  [0]{URL }%
\providecommand \Eprint [0]{\href }%
\providecommand \doibase [0]{https://doi.org/}%
\providecommand \selectlanguage [0]{\@gobble}%
\providecommand \bibinfo  [0]{\@secondoftwo}%
\providecommand \bibfield  [0]{\@secondoftwo}%
\providecommand \translation [1]{[#1]}%
\providecommand \BibitemOpen [0]{}%
\providecommand \bibitemStop [0]{}%
\providecommand \bibitemNoStop [0]{.\EOS\space}%
\providecommand \EOS [0]{\spacefactor3000\relax}%
\providecommand \BibitemShut  [1]{\csname bibitem#1\endcsname}%
\let\auto@bib@innerbib\@empty
\bibitem [{\citenamefont {Anderson}(1973)}]{Anderson1973}%
  \BibitemOpen
  \bibfield  {author} {\bibinfo {author} {\bibfnamefont {P.~W.}\ \bibnamefont
  {Anderson}},\ }\bibfield  {title} {\bibinfo {title} {{Resonating valence
  bonds: A new kind of insulator?}},\ }\href
  {https://doi.org/http://dx.doi.org/10.1016/0025-5408(73)90167-0} {\bibfield
  {journal} {\bibinfo  {journal} {Mater. Res. Bull.}\ }\textbf {\bibinfo
  {volume} {8}},\ \bibinfo {pages} {153} (\bibinfo {year} {1973})}\BibitemShut
  {NoStop}%
\bibitem [{\citenamefont {Balents}(2010)}]{Balents2010}%
  \BibitemOpen
  \bibfield  {author} {\bibinfo {author} {\bibfnamefont {L.}~\bibnamefont
  {Balents}},\ }\bibfield  {title} {\bibinfo {title} {{Spin liquids in
  frustrated magnets.}},\ }\href {https://doi.org/10.1038/nature08917}
  {\bibfield  {journal} {\bibinfo  {journal} {Nature}\ }\textbf {\bibinfo
  {volume} {464}},\ \bibinfo {pages} {199} (\bibinfo {year}
  {2010})}\BibitemShut {NoStop}%
\bibitem [{\citenamefont {Broholm}\ \emph {et~al.}(2020)\citenamefont
  {Broholm}, \citenamefont {Cava}, \citenamefont {Kivelson}, \citenamefont
  {Nocera}, \citenamefont {Norman},\ and\ \citenamefont
  {Senthil}}]{Broholm2020}%
  \BibitemOpen
  \bibfield  {author} {\bibinfo {author} {\bibfnamefont {C.}~\bibnamefont
  {Broholm}}, \bibinfo {author} {\bibfnamefont {R.~J.}\ \bibnamefont {Cava}},
  \bibinfo {author} {\bibfnamefont {S.~A.}\ \bibnamefont {Kivelson}}, \bibinfo
  {author} {\bibfnamefont {D.~G.}\ \bibnamefont {Nocera}}, \bibinfo {author}
  {\bibfnamefont {M.~R.}\ \bibnamefont {Norman}},\ and\ \bibinfo {author}
  {\bibfnamefont {T.}~\bibnamefont {Senthil}},\ }\bibfield  {title} {\bibinfo
  {title} {{Quantum spin liquids}},\ }\href
  {https://doi.org/10.1126/science.aay0668} {\bibfield  {journal} {\bibinfo
  {journal} {Science}\ }\textbf {\bibinfo {volume} {367}},\ \bibinfo {pages}
  {eaay0668} (\bibinfo {year} {2020})}\BibitemShut {NoStop}%
\bibitem [{\citenamefont {Tamura}\ \emph {et~al.}(2006)\citenamefont {Tamura},
  \citenamefont {Nakao},\ and\ \citenamefont {Kato}}]{Tamura2006}%
  \BibitemOpen
  \bibfield  {author} {\bibinfo {author} {\bibfnamefont {M.}~\bibnamefont
  {Tamura}}, \bibinfo {author} {\bibfnamefont {A.}~\bibnamefont {Nakao}},\ and\
  \bibinfo {author} {\bibfnamefont {R.}~\bibnamefont {Kato}},\ }\bibfield
  {title} {\bibinfo {title} {{Frustration-Induced Valence-Bond Ordering in a
  New Quantum Triangular Antiferromagnet Based on [Pd(dmit)2]}},\ }\href
  {https://doi.org/10.1143/JPSJ.75.093701} {\bibfield  {journal} {\bibinfo
  {journal} {J. Phys. Soc. Jpn.}\ }\textbf {\bibinfo {volume} {75}},\ \bibinfo
  {pages} {93701} (\bibinfo {year} {2006})}\BibitemShut {NoStop}%
\bibitem [{\citenamefont {Shimizu}\ \emph {et~al.}(2007)\citenamefont
  {Shimizu}, \citenamefont {Akimoto}, \citenamefont {Tsujii}, \citenamefont
  {Tajima},\ and\ \citenamefont {Kato}}]{Shimizu2007}%
  \BibitemOpen
  \bibfield  {author} {\bibinfo {author} {\bibfnamefont {Y.}~\bibnamefont
  {Shimizu}}, \bibinfo {author} {\bibfnamefont {H.}~\bibnamefont {Akimoto}},
  \bibinfo {author} {\bibfnamefont {H.}~\bibnamefont {Tsujii}}, \bibinfo
  {author} {\bibfnamefont {A.}~\bibnamefont {Tajima}},\ and\ \bibinfo {author}
  {\bibfnamefont {R.}~\bibnamefont {Kato}},\ }\bibfield  {title} {\bibinfo
  {title} {{Mott Transition in a Valence-Bond Solid Insulator with a Triangular
  Lattice}},\ }\href {https://doi.org/10.1103/PhysRevLett.99.256403} {\bibfield
   {journal} {\bibinfo  {journal} {Phys. Rev. Lett.}\ }\textbf {\bibinfo
  {volume} {99}},\ \bibinfo {pages} {256403} (\bibinfo {year}
  {2007})}\BibitemShut {NoStop}%
\bibitem [{\citenamefont {Itou}\ \emph {et~al.}(2009)\citenamefont {Itou},
  \citenamefont {Oyamada}, \citenamefont {Maegawa}, \citenamefont {Kubo},
  \citenamefont {Yamamoto},\ and\ \citenamefont {Kato}}]{Itou2009}%
  \BibitemOpen
  \bibfield  {author} {\bibinfo {author} {\bibfnamefont {T.}~\bibnamefont
  {Itou}}, \bibinfo {author} {\bibfnamefont {A.}~\bibnamefont {Oyamada}},
  \bibinfo {author} {\bibfnamefont {S.}~\bibnamefont {Maegawa}}, \bibinfo
  {author} {\bibfnamefont {K.}~\bibnamefont {Kubo}}, \bibinfo {author}
  {\bibfnamefont {H.~M.}\ \bibnamefont {Yamamoto}},\ and\ \bibinfo {author}
  {\bibfnamefont {R.}~\bibnamefont {Kato}},\ }\bibfield  {title} {\bibinfo
  {title} {{Superconductivity on the border of a spin-gapped Mott insulator:
  NMR studies of the quasi-two-dimensional organic system
  ${\text{EtMe}}_{3}\text{P}{[\text{Pd}{(\text{dmit})}_{2}]}_{2}$}},\ }\href
  {https://doi.org/10.1103/PhysRevB.79.174517} {\bibfield  {journal} {\bibinfo
  {journal} {Phys. Rev. B}\ }\textbf {\bibinfo {volume} {79}},\ \bibinfo
  {pages} {174517} (\bibinfo {year} {2009})}\BibitemShut {NoStop}%
\bibitem [{\citenamefont {Manna}\ \emph {et~al.}(2014)\citenamefont {Manna},
  \citenamefont {de~Souza}, \citenamefont {Kato},\ and\ \citenamefont
  {Lang}}]{Manna2014}%
  \BibitemOpen
  \bibfield  {author} {\bibinfo {author} {\bibfnamefont {R.~S.}\ \bibnamefont
  {Manna}}, \bibinfo {author} {\bibfnamefont {M.}~\bibnamefont {de~Souza}},
  \bibinfo {author} {\bibfnamefont {R.}~\bibnamefont {Kato}},\ and\ \bibinfo
  {author} {\bibfnamefont {M.}~\bibnamefont {Lang}},\ }\bibfield  {title}
  {\bibinfo {title} {{Lattice effects in the quasi-two-dimensional
  valence-bond-solid Mott insulator EtMe${}_{3}$P[Pd(dmit)${}_{2}$]${}_{2}$}},\
  }\href {https://doi.org/10.1103/PhysRevB.89.045113} {\bibfield  {journal}
  {\bibinfo  {journal} {Phys. Rev. B}\ }\textbf {\bibinfo {volume} {89}},\
  \bibinfo {pages} {45113} (\bibinfo {year} {2014})}\BibitemShut {NoStop}%
\bibitem [{\citenamefont {Yoshida}\ \emph {et~al.}(2015)\citenamefont
  {Yoshida}, \citenamefont {Ito}, \citenamefont {Maesato}, \citenamefont
  {Shimizu}, \citenamefont {Hayama}, \citenamefont {Hiramatsu}, \citenamefont
  {Nakamura}, \citenamefont {Kishida}, \citenamefont {Koretsune}, \citenamefont
  {Hotta},\ and\ \citenamefont {Saito}}]{Yoshida2015}%
  \BibitemOpen
  \bibfield  {author} {\bibinfo {author} {\bibfnamefont {Y.}~\bibnamefont
  {Yoshida}}, \bibinfo {author} {\bibfnamefont {H.}~\bibnamefont {Ito}},
  \bibinfo {author} {\bibfnamefont {M.}~\bibnamefont {Maesato}}, \bibinfo
  {author} {\bibfnamefont {Y.}~\bibnamefont {Shimizu}}, \bibinfo {author}
  {\bibfnamefont {H.}~\bibnamefont {Hayama}}, \bibinfo {author} {\bibfnamefont
  {T.}~\bibnamefont {Hiramatsu}}, \bibinfo {author} {\bibfnamefont
  {Y.}~\bibnamefont {Nakamura}}, \bibinfo {author} {\bibfnamefont
  {H.}~\bibnamefont {Kishida}}, \bibinfo {author} {\bibfnamefont
  {T.}~\bibnamefont {Koretsune}}, \bibinfo {author} {\bibfnamefont
  {C.}~\bibnamefont {Hotta}},\ and\ \bibinfo {author} {\bibfnamefont
  {G.}~\bibnamefont {Saito}},\ }\bibfield  {title} {\bibinfo {title}
  {{Spin-disordered quantum phases in a quasi-one-dimensional triangular
  lattice}},\ }\href {https://doi.org/10.1038/nphys3359} {\bibfield  {journal}
  {\bibinfo  {journal} {Nat. Phys.}\ }\textbf {\bibinfo {volume} {11}},\
  \bibinfo {pages} {679} (\bibinfo {year} {2015})}\BibitemShut {NoStop}%
\bibitem [{\citenamefont {Dumm}\ \emph {et~al.}(2000)\citenamefont {Dumm},
  \citenamefont {Loidl}, \citenamefont {Alavi}, \citenamefont {Starkey},
  \citenamefont {Montgomery},\ and\ \citenamefont {Dressel}}]{Dumm2000}%
  \BibitemOpen
  \bibfield  {author} {\bibinfo {author} {\bibfnamefont {M.}~\bibnamefont
  {Dumm}}, \bibinfo {author} {\bibfnamefont {A.}~\bibnamefont {Loidl}},
  \bibinfo {author} {\bibfnamefont {B.}~\bibnamefont {Alavi}}, \bibinfo
  {author} {\bibfnamefont {K.~P.}\ \bibnamefont {Starkey}}, \bibinfo {author}
  {\bibfnamefont {L.~K.}\ \bibnamefont {Montgomery}},\ and\ \bibinfo {author}
  {\bibfnamefont {M.}~\bibnamefont {Dressel}},\ }\bibfield  {title} {\bibinfo
  {title} {{Comprehensive ESR study of the antiferromagnetic ground states in
  the one-dimensional spin systems ${(\mathrm{TMTSF})}_{2}{\mathrm{PF}}_{6},$
  ${(\mathrm{TMTSF})}_{2}{\mathrm{AsF}}_{6},$ and
  ${(\mathrm{TMTTF})}_{2}\mathrm{Br}$}},\ }\href
  {https://link.aps.org/doi/10.1103/PhysRevB.62.6512} {\bibfield  {journal}
  {\bibinfo  {journal} {Phys. Rev. B}\ }\textbf {\bibinfo {volume} {62}},\
  \bibinfo {pages} {6512} (\bibinfo {year} {2000})}\BibitemShut {NoStop}%
\bibitem [{\citenamefont {Hase}\ \emph {et~al.}(1993)\citenamefont {Hase},
  \citenamefont {Terasaki},\ and\ \citenamefont {Uchinokura}}]{Hase1993}%
  \BibitemOpen
  \bibfield  {author} {\bibinfo {author} {\bibfnamefont {M.}~\bibnamefont
  {Hase}}, \bibinfo {author} {\bibfnamefont {I.}~\bibnamefont {Terasaki}},\
  and\ \bibinfo {author} {\bibfnamefont {K.}~\bibnamefont {Uchinokura}},\
  }\bibfield  {title} {\bibinfo {title} {{Observation of the spin-Peierls
  transition in linear ${\mathrm{Cu}}^{2+}$ (spin-1/2) chains in an inorganic
  compound ${\mathrm{CuGeO}}_{3}$}},\ }\href
  {https://doi.org/10.1103/PhysRevLett.70.3651} {\bibfield  {journal} {\bibinfo
   {journal} {Phys. Rev. Lett.}\ }\textbf {\bibinfo {volume} {70}},\ \bibinfo
  {pages} {3651} (\bibinfo {year} {1993})}\BibitemShut {NoStop}%
\bibitem [{\citenamefont {Hermanns}\ \emph {et~al.}(2015)\citenamefont
  {Hermanns}, \citenamefont {Trebst},\ and\ \citenamefont
  {Rosch}}]{Hermanns2015}%
  \BibitemOpen
  \bibfield  {author} {\bibinfo {author} {\bibfnamefont {M.}~\bibnamefont
  {Hermanns}}, \bibinfo {author} {\bibfnamefont {S.}~\bibnamefont {Trebst}},\
  and\ \bibinfo {author} {\bibfnamefont {A.}~\bibnamefont {Rosch}},\ }\bibfield
   {title} {\bibinfo {title} {{Spin-Peierls Instability of Three-Dimensional
  Spin Liquids with Majorana Fermi Surfaces}},\ }\href
  {https://doi.org/10.1103/PhysRevLett.115.177205} {\bibfield  {journal}
  {\bibinfo  {journal} {Phys. Rev. Lett.}\ }\textbf {\bibinfo {volume} {115}},\
  \bibinfo {pages} {177205} (\bibinfo {year} {2015})}\BibitemShut {NoStop}%
\bibitem [{\citenamefont {Norman}\ \emph {et~al.}(2020)\citenamefont {Norman},
  \citenamefont {Laurita},\ and\ \citenamefont {Hsieh}}]{Norman2020}%
  \BibitemOpen
  \bibfield  {author} {\bibinfo {author} {\bibfnamefont {M.~R.}\ \bibnamefont
  {Norman}}, \bibinfo {author} {\bibfnamefont {N.~J.}\ \bibnamefont
  {Laurita}},\ and\ \bibinfo {author} {\bibfnamefont {D.}~\bibnamefont
  {Hsieh}},\ }\bibfield  {title} {\bibinfo {title} {{Valence bond phases of
  herbertsmithite and related copper kagome materials}},\ }\href
  {https://doi.org/10.1103/PhysRevResearch.2.013055} {\bibfield  {journal}
  {\bibinfo  {journal} {Phys. Rev. Research}\ }\textbf {\bibinfo {volume}
  {2}},\ \bibinfo {pages} {13055} (\bibinfo {year} {2020})}\BibitemShut
  {NoStop}%
\bibitem [{\citenamefont {Kimchi}\ \emph {et~al.}(2018)\citenamefont {Kimchi},
  \citenamefont {Nahum},\ and\ \citenamefont {Senthil}}]{Kimchi2018}%
  \BibitemOpen
  \bibfield  {author} {\bibinfo {author} {\bibfnamefont {I.}~\bibnamefont
  {Kimchi}}, \bibinfo {author} {\bibfnamefont {A.}~\bibnamefont {Nahum}},\ and\
  \bibinfo {author} {\bibfnamefont {T.}~\bibnamefont {Senthil}},\ }\bibfield
  {title} {\bibinfo {title} {{Valence Bonds in Random Quantum Magnets: Theory
  and Application to ${\mathrm{YbMgGaO}}_{4}$}},\ }\href
  {https://doi.org/10.1103/PhysRevX.8.031028} {\bibfield  {journal} {\bibinfo
  {journal} {Phys. Rev. X}\ }\textbf {\bibinfo {volume} {8}},\ \bibinfo {pages}
  {31028} (\bibinfo {year} {2018})}\BibitemShut {NoStop}%
\bibitem [{\citenamefont {Riedl}\ \emph {et~al.}(2019)\citenamefont {Riedl},
  \citenamefont {Valent{\'{i}}},\ and\ \citenamefont {Winter}}]{Riedl2019}%
  \BibitemOpen
  \bibfield  {author} {\bibinfo {author} {\bibfnamefont {K.}~\bibnamefont
  {Riedl}}, \bibinfo {author} {\bibfnamefont {R.}~\bibnamefont
  {Valent{\'{i}}}},\ and\ \bibinfo {author} {\bibfnamefont {S.~M.}\
  \bibnamefont {Winter}},\ }\bibfield  {title} {\bibinfo {title} {{Critical
  spin liquid versus valence-bond glass in a triangular-lattice organic
  antiferromagnet}},\ }\href {https://doi.org/10.1038/s41467-019-10604-3}
  {\bibfield  {journal} {\bibinfo  {journal} {Nat. Commun.}\ }\textbf {\bibinfo
  {volume} {10}},\ \bibinfo {pages} {2561} (\bibinfo {year}
  {2019})}\BibitemShut {NoStop}%
\bibitem [{\citenamefont {Miksch}\ \emph {et~al.}(2021)\citenamefont {Miksch},
  \citenamefont {Pustogow}, \citenamefont {{Javaheri Rahim}}, \citenamefont
  {Bardin}, \citenamefont {Kanoda}, \citenamefont {Schlueter}, \citenamefont
  {H{\"{u}}bner}, \citenamefont {Scheffler},\ and\ \citenamefont
  {Dressel}}]{Miksch2021}%
  \BibitemOpen
  \bibfield  {author} {\bibinfo {author} {\bibfnamefont {B.}~\bibnamefont
  {Miksch}}, \bibinfo {author} {\bibfnamefont {A.}~\bibnamefont {Pustogow}},
  \bibinfo {author} {\bibfnamefont {M.}~\bibnamefont {{Javaheri Rahim}}},
  \bibinfo {author} {\bibfnamefont {A.~A.}\ \bibnamefont {Bardin}}, \bibinfo
  {author} {\bibfnamefont {K.}~\bibnamefont {Kanoda}}, \bibinfo {author}
  {\bibfnamefont {J.~A.}\ \bibnamefont {Schlueter}}, \bibinfo {author}
  {\bibfnamefont {R.}~\bibnamefont {H{\"{u}}bner}}, \bibinfo {author}
  {\bibfnamefont {M.}~\bibnamefont {Scheffler}},\ and\ \bibinfo {author}
  {\bibfnamefont {M.}~\bibnamefont {Dressel}},\ }\bibfield  {title} {\bibinfo
  {title} {{Gapped magnetic ground state in quantum spin liquid candidate
  $\kappa$-(BEDT-TTF)$_2$Cu$_2$(CN)$_3$}},\ }\href
  {https://doi.org/10.1126/science.abc6363} {\bibfield  {journal} {\bibinfo
  {journal} {Science}\ }\textbf {\bibinfo {volume} {372}},\ \bibinfo {pages}
  {276 LP } (\bibinfo {year} {2021})}\BibitemShut {NoStop}%
\bibitem [{\citenamefont {Pustogow}(2022)}]{Pustogow2022}%
  \BibitemOpen
  \bibfield  {author} {\bibinfo {author} {\bibfnamefont {A.}~\bibnamefont
  {Pustogow}},\ }\href {https://doi.org/10.3390/solids3010007} {\bibinfo
  {title} {{Thirty-Year Anniversary of $\kappa$-(BEDT-TTF)$_2$Cu$_2$(CN)$_3$:
  Reconciling the Spin Gap in a Spin-Liquid Candidate}}} (\bibinfo {year}
  {2022})\BibitemShut {NoStop}%
\bibitem [{\citenamefont {Yamashita}\ \emph {et~al.}(2008)\citenamefont
  {Yamashita}, \citenamefont {Nakazawa}, \citenamefont {Oguni}, \citenamefont
  {Oshima}, \citenamefont {Nojiri}, \citenamefont {Shimizu}, \citenamefont
  {Miyagawa},\ and\ \citenamefont {Kanoda}}]{Yamashita2008}%
  \BibitemOpen
  \bibfield  {author} {\bibinfo {author} {\bibfnamefont {S.}~\bibnamefont
  {Yamashita}}, \bibinfo {author} {\bibfnamefont {Y.}~\bibnamefont {Nakazawa}},
  \bibinfo {author} {\bibfnamefont {M.}~\bibnamefont {Oguni}}, \bibinfo
  {author} {\bibfnamefont {Y.}~\bibnamefont {Oshima}}, \bibinfo {author}
  {\bibfnamefont {H.}~\bibnamefont {Nojiri}}, \bibinfo {author} {\bibfnamefont
  {Y.}~\bibnamefont {Shimizu}}, \bibinfo {author} {\bibfnamefont
  {K.}~\bibnamefont {Miyagawa}},\ and\ \bibinfo {author} {\bibfnamefont
  {K.}~\bibnamefont {Kanoda}},\ }\bibfield  {title} {\bibinfo {title}
  {{Thermodynamic properties of a spin-1/2 spin-liquid state in a [kappa]-type
  organic salt}},\ }\href {http://dx.doi.org/10.1038/nphys942} {\bibfield
  {journal} {\bibinfo  {journal} {Nat. Phys.}\ }\textbf {\bibinfo {volume}
  {4}},\ \bibinfo {pages} {459} (\bibinfo {year} {2008})}\BibitemShut {NoStop}%
\bibitem [{\citenamefont {Yamashita}\ \emph {et~al.}(2009)\citenamefont
  {Yamashita}, \citenamefont {Nakata}, \citenamefont {Kasahara}, \citenamefont
  {Sasaki}, \citenamefont {Yoneyama}, \citenamefont {Kobayashi}, \citenamefont
  {Fujimoto}, \citenamefont {Shibauchi},\ and\ \citenamefont
  {Matsuda}}]{Yamashita2009}%
  \BibitemOpen
  \bibfield  {author} {\bibinfo {author} {\bibfnamefont {M.}~\bibnamefont
  {Yamashita}}, \bibinfo {author} {\bibfnamefont {N.}~\bibnamefont {Nakata}},
  \bibinfo {author} {\bibfnamefont {Y.}~\bibnamefont {Kasahara}}, \bibinfo
  {author} {\bibfnamefont {T.}~\bibnamefont {Sasaki}}, \bibinfo {author}
  {\bibfnamefont {N.}~\bibnamefont {Yoneyama}}, \bibinfo {author}
  {\bibfnamefont {N.}~\bibnamefont {Kobayashi}}, \bibinfo {author}
  {\bibfnamefont {S.}~\bibnamefont {Fujimoto}}, \bibinfo {author}
  {\bibfnamefont {T.}~\bibnamefont {Shibauchi}},\ and\ \bibinfo {author}
  {\bibfnamefont {Y.}~\bibnamefont {Matsuda}},\ }\bibfield  {title} {\bibinfo
  {title} {{Thermal-transport measurements in a quantum spin-liquid state of
  the frustrated triangular magnet
  nphys1134-m6gif1601313-(BEDT-TTF)2Cu2(CN)3}},\ }\href
  {http://dx.doi.org/10.1038/nphys1134
  http://www.nature.com/nphys/journal/v5/n1/suppinfo/nphys1134_S1.html}
  {\bibfield  {journal} {\bibinfo  {journal} {Nat. Phys.}\ }\textbf {\bibinfo
  {volume} {5}},\ \bibinfo {pages} {44} (\bibinfo {year} {2009})}\BibitemShut
  {NoStop}%
\bibitem [{\citenamefont {Lefebvre}\ \emph {et~al.}(2000)\citenamefont
  {Lefebvre}, \citenamefont {Wzietek}, \citenamefont {Brown}, \citenamefont
  {Bourbonnais}, \citenamefont {J{\'{e}}rome}, \citenamefont
  {M{\'{e}}zi{\`{e}}re}, \citenamefont {Fourmigu{\'{e}}},\ and\ \citenamefont
  {Batail}}]{Lefebvre2000}%
  \BibitemOpen
  \bibfield  {author} {\bibinfo {author} {\bibfnamefont {S.}~\bibnamefont
  {Lefebvre}}, \bibinfo {author} {\bibfnamefont {P.}~\bibnamefont {Wzietek}},
  \bibinfo {author} {\bibfnamefont {S.}~\bibnamefont {Brown}}, \bibinfo
  {author} {\bibfnamefont {C.}~\bibnamefont {Bourbonnais}}, \bibinfo {author}
  {\bibfnamefont {D.}~\bibnamefont {J{\'{e}}rome}}, \bibinfo {author}
  {\bibfnamefont {C.}~\bibnamefont {M{\'{e}}zi{\`{e}}re}}, \bibinfo {author}
  {\bibfnamefont {M.}~\bibnamefont {Fourmigu{\'{e}}}},\ and\ \bibinfo {author}
  {\bibfnamefont {P.}~\bibnamefont {Batail}},\ }\bibfield  {title} {\bibinfo
  {title} {{Mott transition, antiferromagnetism, and unconventional
  superconductivity in layered organic superconductors.}},\ }\href@noop {}
  {\bibfield  {journal} {\bibinfo  {journal} {Phys. Rev. Lett.}\ }\textbf
  {\bibinfo {volume} {85}},\ \bibinfo {pages} {5420} (\bibinfo {year}
  {2000})}\BibitemShut {NoStop}%
\bibitem [{\citenamefont {Limelette}\ \emph {et~al.}(2003)\citenamefont
  {Limelette}, \citenamefont {Wzietek}, \citenamefont {Florens}, \citenamefont
  {Georges}, \citenamefont {Costi}, \citenamefont {Pasquier}, \citenamefont
  {J{\'{e}}rome}, \citenamefont {M{\'{e}}zi{\`{e}}re},\ and\ \citenamefont
  {Batail}}]{Limelette2003a}%
  \BibitemOpen
  \bibfield  {author} {\bibinfo {author} {\bibfnamefont {P.}~\bibnamefont
  {Limelette}}, \bibinfo {author} {\bibfnamefont {P.}~\bibnamefont {Wzietek}},
  \bibinfo {author} {\bibfnamefont {S.}~\bibnamefont {Florens}}, \bibinfo
  {author} {\bibfnamefont {A.}~\bibnamefont {Georges}}, \bibinfo {author}
  {\bibfnamefont {T.~A.}\ \bibnamefont {Costi}}, \bibinfo {author}
  {\bibfnamefont {C.}~\bibnamefont {Pasquier}}, \bibinfo {author}
  {\bibfnamefont {D.}~\bibnamefont {J{\'{e}}rome}}, \bibinfo {author}
  {\bibfnamefont {C.}~\bibnamefont {M{\'{e}}zi{\`{e}}re}},\ and\ \bibinfo
  {author} {\bibfnamefont {P.}~\bibnamefont {Batail}},\ }\bibfield  {title}
  {\bibinfo {title} {{Mott Transition and Transport Crossovers in the Organic
  Compound $\kappa$-(BEDT-TTF)$_2$Cu[N(CN)$_2$]Cl}},\ }\href
  {https://link.aps.org/doi/10.1103/PhysRevLett.91.016401} {\bibfield
  {journal} {\bibinfo  {journal} {Phys. Rev. Lett.}\ }\textbf {\bibinfo
  {volume} {91}},\ \bibinfo {pages} {16401} (\bibinfo {year}
  {2003})}\BibitemShut {NoStop}%
\bibitem [{\citenamefont {Kurosaki}\ \emph {et~al.}(2005)\citenamefont
  {Kurosaki}, \citenamefont {Shimizu}, \citenamefont {Miyagawa}, \citenamefont
  {Kanoda},\ and\ \citenamefont {Saito}}]{Kurosaki2005}%
  \BibitemOpen
  \bibfield  {author} {\bibinfo {author} {\bibfnamefont {Y.}~\bibnamefont
  {Kurosaki}}, \bibinfo {author} {\bibfnamefont {Y.}~\bibnamefont {Shimizu}},
  \bibinfo {author} {\bibfnamefont {K.}~\bibnamefont {Miyagawa}}, \bibinfo
  {author} {\bibfnamefont {K.}~\bibnamefont {Kanoda}},\ and\ \bibinfo {author}
  {\bibfnamefont {G.}~\bibnamefont {Saito}},\ }\bibfield  {title} {\bibinfo
  {title} {{Mott Transition from a Spin Liquid to a Fermi Liquid in the
  Spin-Frustrated Organic Conductor
  $\ensuremath{\kappa}\mathrm{\text{\ensuremath{-}}}(\mathrm{ET}{)}_{2}{\mathrm{Cu}}_{2}(\mathrm{CN}{)}_{3}$}},\
  }\href {https://link.aps.org/doi/10.1103/PhysRevLett.95.177001} {\bibfield
  {journal} {\bibinfo  {journal} {Phys. Rev. Lett.}\ }\textbf {\bibinfo
  {volume} {95}},\ \bibinfo {pages} {177001} (\bibinfo {year}
  {2005})}\BibitemShut {NoStop}%
\bibitem [{\citenamefont {Shimizu}\ \emph {et~al.}(2016)\citenamefont
  {Shimizu}, \citenamefont {Hiramatsu}, \citenamefont {Maesato}, \citenamefont
  {Otsuka}, \citenamefont {Yamochi}, \citenamefont {Ono}, \citenamefont {Itoh},
  \citenamefont {Yoshida}, \citenamefont {Takigawa}, \citenamefont {Yoshida},\
  and\ \citenamefont {Saito}}]{Shimizu2016}%
  \BibitemOpen
  \bibfield  {author} {\bibinfo {author} {\bibfnamefont {Y.}~\bibnamefont
  {Shimizu}}, \bibinfo {author} {\bibfnamefont {T.}~\bibnamefont {Hiramatsu}},
  \bibinfo {author} {\bibfnamefont {M.}~\bibnamefont {Maesato}}, \bibinfo
  {author} {\bibfnamefont {A.}~\bibnamefont {Otsuka}}, \bibinfo {author}
  {\bibfnamefont {H.}~\bibnamefont {Yamochi}}, \bibinfo {author} {\bibfnamefont
  {A.}~\bibnamefont {Ono}}, \bibinfo {author} {\bibfnamefont {M.}~\bibnamefont
  {Itoh}}, \bibinfo {author} {\bibfnamefont {M.}~\bibnamefont {Yoshida}},
  \bibinfo {author} {\bibfnamefont {M.}~\bibnamefont {Takigawa}}, \bibinfo
  {author} {\bibfnamefont {Y.}~\bibnamefont {Yoshida}},\ and\ \bibinfo {author}
  {\bibfnamefont {G.}~\bibnamefont {Saito}},\ }\bibfield  {title} {\bibinfo
  {title} {{Pressure-Tuned Exchange Coupling of a Quantum Spin Liquid in the
  Molecular Triangular Lattice
  $\ensuremath{\kappa}\text{\ensuremath{-}}(\mathrm{ET}{)}_{2}{\mathrm{Ag}}_{2}(\mathrm{CN}{)}_{3}$}},\
  }\href {https://link.aps.org/doi/10.1103/PhysRevLett.117.107203} {\bibfield
  {journal} {\bibinfo  {journal} {Phys. Rev. Lett.}\ }\textbf {\bibinfo
  {volume} {117}},\ \bibinfo {pages} {107203} (\bibinfo {year}
  {2016})}\BibitemShut {NoStop}%
\bibitem [{\citenamefont {Furukawa}\ \emph {et~al.}(2018)\citenamefont
  {Furukawa}, \citenamefont {Kobashi}, \citenamefont {Kurosaki}, \citenamefont
  {Miyagawa},\ and\ \citenamefont {Kanoda}}]{Furukawa2018}%
  \BibitemOpen
  \bibfield  {author} {\bibinfo {author} {\bibfnamefont {T.}~\bibnamefont
  {Furukawa}}, \bibinfo {author} {\bibfnamefont {K.}~\bibnamefont {Kobashi}},
  \bibinfo {author} {\bibfnamefont {Y.}~\bibnamefont {Kurosaki}}, \bibinfo
  {author} {\bibfnamefont {K.}~\bibnamefont {Miyagawa}},\ and\ \bibinfo
  {author} {\bibfnamefont {K.}~\bibnamefont {Kanoda}},\ }\bibfield  {title}
  {\bibinfo {title} {{Quasi-continuous transition from a Fermi liquid to a spin
  liquid in $\kappa$-(ET)2Cu2(CN)3}},\ }\href
  {https://doi.org/10.1038/s41467-017-02679-7} {\bibfield  {journal} {\bibinfo
  {journal} {Nat. Commun.}\ }\textbf {\bibinfo {volume} {9}},\ \bibinfo {pages}
  {307} (\bibinfo {year} {2018})}\BibitemShut {NoStop}%
\bibitem [{\citenamefont {Pustogow}\ \emph {et~al.}(2018)\citenamefont
  {Pustogow}, \citenamefont {Bories}, \citenamefont {L{\"{o}}hle},
  \citenamefont {R{\"{o}}sslhuber}, \citenamefont {Zhukova}, \citenamefont
  {Gorshunov}, \citenamefont {Tomi{\'{c}}}, \citenamefont {Schlueter},
  \citenamefont {H{\"{u}}bner}, \citenamefont {Hiramatsu}, \citenamefont
  {Yoshida}, \citenamefont {Saito}, \citenamefont {Kato}, \citenamefont {Lee},
  \citenamefont {Dobrosavljevi{\'{c}}}, \citenamefont {Fratini},\ and\
  \citenamefont {Dressel}}]{Pustogow2018}%
  \BibitemOpen
  \bibfield  {author} {\bibinfo {author} {\bibfnamefont {A.}~\bibnamefont
  {Pustogow}}, \bibinfo {author} {\bibfnamefont {M.}~\bibnamefont {Bories}},
  \bibinfo {author} {\bibfnamefont {A.}~\bibnamefont {L{\"{o}}hle}}, \bibinfo
  {author} {\bibfnamefont {R.}~\bibnamefont {R{\"{o}}sslhuber}}, \bibinfo
  {author} {\bibfnamefont {E.}~\bibnamefont {Zhukova}}, \bibinfo {author}
  {\bibfnamefont {B.}~\bibnamefont {Gorshunov}}, \bibinfo {author}
  {\bibfnamefont {S.}~\bibnamefont {Tomi{\'{c}}}}, \bibinfo {author}
  {\bibfnamefont {J.~A.}\ \bibnamefont {Schlueter}}, \bibinfo {author}
  {\bibfnamefont {R.}~\bibnamefont {H{\"{u}}bner}}, \bibinfo {author}
  {\bibfnamefont {T.}~\bibnamefont {Hiramatsu}}, \bibinfo {author}
  {\bibfnamefont {Y.}~\bibnamefont {Yoshida}}, \bibinfo {author} {\bibfnamefont
  {G.}~\bibnamefont {Saito}}, \bibinfo {author} {\bibfnamefont
  {R.}~\bibnamefont {Kato}}, \bibinfo {author} {\bibfnamefont {T.-H.}\
  \bibnamefont {Lee}}, \bibinfo {author} {\bibfnamefont {V.}~\bibnamefont
  {Dobrosavljevi{\'{c}}}}, \bibinfo {author} {\bibfnamefont {S.}~\bibnamefont
  {Fratini}},\ and\ \bibinfo {author} {\bibfnamefont {M.}~\bibnamefont
  {Dressel}},\ }\bibfield  {title} {\bibinfo {title} {{Quantum spin liquids
  unveil the genuine Mott state}},\ }\href
  {https://doi.org/10.1038/s41563-018-0140-3} {\bibfield  {journal} {\bibinfo
  {journal} {Nat. Mater.}\ }\textbf {\bibinfo {volume} {17}},\ \bibinfo {pages}
  {773} (\bibinfo {year} {2018})}\BibitemShut {NoStop}%
\bibitem [{\citenamefont {Manna}\ \emph {et~al.}(2010)\citenamefont {Manna},
  \citenamefont {de~Souza}, \citenamefont {Br{\"{u}}hl}, \citenamefont
  {Schlueter},\ and\ \citenamefont {Lang}}]{Manna2010}%
  \BibitemOpen
  \bibfield  {author} {\bibinfo {author} {\bibfnamefont {R.~S.}\ \bibnamefont
  {Manna}}, \bibinfo {author} {\bibfnamefont {M.}~\bibnamefont {de~Souza}},
  \bibinfo {author} {\bibfnamefont {A.}~\bibnamefont {Br{\"{u}}hl}}, \bibinfo
  {author} {\bibfnamefont {J.~A.}\ \bibnamefont {Schlueter}},\ and\ \bibinfo
  {author} {\bibfnamefont {M.}~\bibnamefont {Lang}},\ }\bibfield  {title}
  {\bibinfo {title} {{Lattice Effects and Entropy Release at the
  Low-Temperature Phase Transition in the Spin-Liquid Candidate
  $\ensuremath{\kappa}\mathrm{\text{\ensuremath{-}}}(\mathrm{BEDT}\mathrm{\text{\ensuremath{-}}}\mathrm{TTF}{)}_{2}{\mathrm{Cu}}_{2}(\mathrm{CN}{)}_{3}$}},\
  }\href {https://doi.org/10.1103/PhysRevLett.104.016403} {\bibfield  {journal}
  {\bibinfo  {journal} {Phys. Rev. Lett.}\ }\textbf {\bibinfo {volume} {104}},\
  \bibinfo {pages} {16403} (\bibinfo {year} {2010})}\BibitemShut {NoStop}%
\bibitem [{\citenamefont {Shimizu}\ \emph {et~al.}(2003)\citenamefont
  {Shimizu}, \citenamefont {Miyagawa}, \citenamefont {Kanoda}, \citenamefont
  {Maesato},\ and\ \citenamefont {Saito}}]{Shimizu2003}%
  \BibitemOpen
  \bibfield  {author} {\bibinfo {author} {\bibfnamefont {Y.}~\bibnamefont
  {Shimizu}}, \bibinfo {author} {\bibfnamefont {K.}~\bibnamefont {Miyagawa}},
  \bibinfo {author} {\bibfnamefont {K.}~\bibnamefont {Kanoda}}, \bibinfo
  {author} {\bibfnamefont {M.}~\bibnamefont {Maesato}},\ and\ \bibinfo {author}
  {\bibfnamefont {G.}~\bibnamefont {Saito}},\ }\bibfield  {title} {\bibinfo
  {title} {{Spin Liquid State in an Organic Mott Insulator with a Triangular
  Lattice}},\ }\href {https://link.aps.org/doi/10.1103/PhysRevLett.91.107001}
  {\bibfield  {journal} {\bibinfo  {journal} {Phys. Rev. Lett.}\ }\textbf
  {\bibinfo {volume} {91}},\ \bibinfo {pages} {107001} (\bibinfo {year}
  {2003})}\BibitemShut {NoStop}%
\bibitem [{\citenamefont {Furukawa}\ \emph {et~al.}(2015)\citenamefont
  {Furukawa}, \citenamefont {Miyagawa}, \citenamefont {Taniguchi},
  \citenamefont {Kato},\ and\ \citenamefont {Kanoda}}]{Furukawa2015}%
  \BibitemOpen
  \bibfield  {author} {\bibinfo {author} {\bibfnamefont {T.}~\bibnamefont
  {Furukawa}}, \bibinfo {author} {\bibfnamefont {K.}~\bibnamefont {Miyagawa}},
  \bibinfo {author} {\bibfnamefont {H.}~\bibnamefont {Taniguchi}}, \bibinfo
  {author} {\bibfnamefont {R.}~\bibnamefont {Kato}},\ and\ \bibinfo {author}
  {\bibfnamefont {K.}~\bibnamefont {Kanoda}},\ }\bibfield  {title} {\bibinfo
  {title} {{Quantum criticality of Mott transition in organic materials}},\
  }\href {http://dx.doi.org/10.1038/nphys3235 http://10.0.4.14/nphys3235
  http://www.nature.com/nphys/journal/v11/n3/abs/nphys3235.html#supplementary-information}
  {\bibfield  {journal} {\bibinfo  {journal} {Nat. Phys.}\ }\textbf {\bibinfo
  {volume} {11}},\ \bibinfo {pages} {221} (\bibinfo {year} {2015})}\BibitemShut
  {NoStop}%
\bibitem [{\citenamefont {Terletska}\ \emph {et~al.}(2011)\citenamefont
  {Terletska}, \citenamefont {Vu{\v{c}}i{\v{c}}evi{\'{c}}}, \citenamefont
  {Tanaskovi{\'{c}}},\ and\ \citenamefont
  {Dobrosavljevi{\'{c}}}}]{Terletska2011}%
  \BibitemOpen
  \bibfield  {author} {\bibinfo {author} {\bibfnamefont {H.}~\bibnamefont
  {Terletska}}, \bibinfo {author} {\bibfnamefont {J.}~\bibnamefont
  {Vu{\v{c}}i{\v{c}}evi{\'{c}}}}, \bibinfo {author} {\bibfnamefont
  {D.}~\bibnamefont {Tanaskovi{\'{c}}}},\ and\ \bibinfo {author} {\bibfnamefont
  {V.}~\bibnamefont {Dobrosavljevi{\'{c}}}},\ }\bibfield  {title} {\bibinfo
  {title} {{Quantum Critical Transport near the Mott Transition}},\ }\href
  {https://link.aps.org/doi/10.1103/PhysRevLett.107.026401} {\bibfield
  {journal} {\bibinfo  {journal} {Phys. Rev. Lett.}\ }\textbf {\bibinfo
  {volume} {107}},\ \bibinfo {pages} {26401} (\bibinfo {year}
  {2011})}\BibitemShut {NoStop}%
\bibitem [{\citenamefont {Vu{\v{c}}i{\v{c}}evi{\'{c}}}\ \emph
  {et~al.}(2013)\citenamefont {Vu{\v{c}}i{\v{c}}evi{\'{c}}}, \citenamefont
  {Terletska}, \citenamefont {Tanaskovi{\'{c}}},\ and\ \citenamefont
  {Dobrosavljevi{\'{c}}}}]{Vucicevic2013}%
  \BibitemOpen
  \bibfield  {author} {\bibinfo {author} {\bibfnamefont {J.}~\bibnamefont
  {Vu{\v{c}}i{\v{c}}evi{\'{c}}}}, \bibinfo {author} {\bibfnamefont
  {H.}~\bibnamefont {Terletska}}, \bibinfo {author} {\bibfnamefont
  {D.}~\bibnamefont {Tanaskovi{\'{c}}}},\ and\ \bibinfo {author} {\bibfnamefont
  {V.}~\bibnamefont {Dobrosavljevi{\'{c}}}},\ }\bibfield  {title} {\bibinfo
  {title} {{Finite-temperature crossover and the quantum Widom line near the
  Mott transition}},\ }\href
  {https://link.aps.org/doi/10.1103/PhysRevB.88.075143} {\bibfield  {journal}
  {\bibinfo  {journal} {Phys. Rev. B}\ }\textbf {\bibinfo {volume} {88}},\
  \bibinfo {pages} {75143} (\bibinfo {year} {2013})}\BibitemShut {NoStop}%
\bibitem [{\citenamefont {Pustogow}\ \emph
  {et~al.}(2021{\natexlab{a}})\citenamefont {Pustogow}, \citenamefont {Saito},
  \citenamefont {L{\"{o}}hle}, \citenamefont {{Sanz Alonso}}, \citenamefont
  {Kawamoto}, \citenamefont {Dobrosavljevi´c}, \citenamefont {Dressel},\ and\
  \citenamefont {Fratini}}]{Pustogow2021-Landau}%
  \BibitemOpen
  \bibfield  {author} {\bibinfo {author} {\bibfnamefont {A.}~\bibnamefont
  {Pustogow}}, \bibinfo {author} {\bibfnamefont {Y.}~\bibnamefont {Saito}},
  \bibinfo {author} {\bibfnamefont {A.}~\bibnamefont {L{\"{o}}hle}}, \bibinfo
  {author} {\bibfnamefont {M.}~\bibnamefont {{Sanz Alonso}}}, \bibinfo {author}
  {\bibfnamefont {A.}~\bibnamefont {Kawamoto}}, \bibinfo {author}
  {\bibfnamefont {V.}~\bibnamefont {Dobrosavljevi´c}}, \bibinfo {author}
  {\bibfnamefont {M.}~\bibnamefont {Dressel}},\ and\ \bibinfo {author}
  {\bibfnamefont {S.}~\bibnamefont {Fratini}},\ }\bibfield  {title} {\bibinfo
  {title} {{Rise and fall of Landau's quasiparticles while approaching the Mott
  transition}},\ }\href {https://doi.org/10.1038/s41467-021-21741-z} {\bibfield
   {journal} {\bibinfo  {journal} {Nat. Commun.}\ }\textbf {\bibinfo {volume}
  {12}},\ \bibinfo {pages} {1571} (\bibinfo {year} {2021}{\natexlab{a}})} \BibitemShut
  {NoStop}%
\bibitem [{\citenamefont {Suda}\ \emph {et~al.}(2014)\citenamefont {Suda},
  \citenamefont {Kawasugi}, \citenamefont {Minari}, \citenamefont {Tsukagoshi},
  \citenamefont {Kato},\ and\ \citenamefont {Yamamoto}}]{Suda2014}%
  \BibitemOpen
  \bibfield  {author} {\bibinfo {author} {\bibfnamefont {M.}~\bibnamefont
  {Suda}}, \bibinfo {author} {\bibfnamefont {Y.}~\bibnamefont {Kawasugi}},
  \bibinfo {author} {\bibfnamefont {T.}~\bibnamefont {Minari}}, \bibinfo
  {author} {\bibfnamefont {K.}~\bibnamefont {Tsukagoshi}}, \bibinfo {author}
  {\bibfnamefont {R.}~\bibnamefont {Kato}},\ and\ \bibinfo {author}
  {\bibfnamefont {H.~M.}\ \bibnamefont {Yamamoto}},\ }\bibfield  {title}
  {{\bibinfo {title} {{Strain-tunable superconducting
  field-effect transistor with an organic strongly-correlated electron
  system.}}},\ }\href {https://doi.org/10.1002/adma.201305797} {\bibfield
  {journal} {\bibinfo  {journal} {Adv. Mater.}\ }\textbf {\bibinfo {volume}
  {26}},\ \bibinfo {pages} {3490} (\bibinfo {year} {2014})}\BibitemShut
  {NoStop}%
\bibitem [{\citenamefont {Shimizu}\ \emph {et~al.}(2011)\citenamefont
  {Shimizu}, \citenamefont {Maesato},\ and\ \citenamefont
  {Saito}}]{Shimizu2011}%
  \BibitemOpen
  \bibfield  {author} {\bibinfo {author} {\bibfnamefont {Y.}~\bibnamefont
  {Shimizu}}, \bibinfo {author} {\bibfnamefont {M.}~\bibnamefont {Maesato}},\
  and\ \bibinfo {author} {\bibfnamefont {G.}~\bibnamefont {Saito}},\ }\bibfield
   {title} {\bibinfo {title} {{Uniaxial Strain Effects on Mott and
  Superconducting Transitions in $\kappa$-(ET)2Cu2(CN)3}},\ }\href
  {https://doi.org/10.1143/JPSJ.80.074702} {\bibfield  {journal} {\bibinfo
  {journal} {J. Phys. Soc. Jpn.}\ }\textbf {\bibinfo {volume} {80}},\ \bibinfo
  {pages} {74702} (\bibinfo {year} {2011})}\BibitemShut {NoStop}%
\bibitem [{\citenamefont {Saito}\ \emph {et~al.}(2018)\citenamefont {Saito},
  \citenamefont {Minamidate}, \citenamefont {Kawamoto}, \citenamefont
  {Matsunaga},\ and\ \citenamefont {Nomura}}]{Saito2018}%
  \BibitemOpen
  \bibfield  {author} {\bibinfo {author} {\bibfnamefont {Y.}~\bibnamefont
  {Saito}}, \bibinfo {author} {\bibfnamefont {T.}~\bibnamefont {Minamidate}},
  \bibinfo {author} {\bibfnamefont {A.}~\bibnamefont {Kawamoto}}, \bibinfo
  {author} {\bibfnamefont {N.}~\bibnamefont {Matsunaga}},\ and\ \bibinfo
  {author} {\bibfnamefont {K.}~\bibnamefont {Nomura}},\ }\bibfield  {title}
  {\bibinfo {title} {{Site-specific $^{13}$C NMR study on the locally distorted
  triangular lattice of the organic conductor
  $\kappa$-(BEDT-TTF)$_2$Cu$_2$(CN)$_3$}},\ }\href
  {https://doi.org/10.1103/PhysRevB.98.205141} {\bibfield  {journal} {\bibinfo
  {journal} {Phys. Rev. B}\ }\textbf {\bibinfo {volume} {98}},\ \bibinfo
  {pages} {205141} (\bibinfo {year} {2018})}\BibitemShut {NoStop}%
\bibitem [{\citenamefont {Itoh}\ \emph {et~al.}(2013)\citenamefont {Itoh},
  \citenamefont {Itoh}, \citenamefont {Naka}, \citenamefont {Saito},
  \citenamefont {Hosako}, \citenamefont {Yoneyama}, \citenamefont {Ishihara},
  \citenamefont {Sasaki},\ and\ \citenamefont {Iwai}}]{Itoh2013}%
  \BibitemOpen
  \bibfield  {author} {\bibinfo {author} {\bibfnamefont {K.}~\bibnamefont
  {Itoh}}, \bibinfo {author} {\bibfnamefont {H.}~\bibnamefont {Itoh}}, \bibinfo
  {author} {\bibfnamefont {M.}~\bibnamefont {Naka}}, \bibinfo {author}
  {\bibfnamefont {S.}~\bibnamefont {Saito}}, \bibinfo {author} {\bibfnamefont
  {I.}~\bibnamefont {Hosako}}, \bibinfo {author} {\bibfnamefont
  {N.}~\bibnamefont {Yoneyama}}, \bibinfo {author} {\bibfnamefont
  {S.}~\bibnamefont {Ishihara}}, \bibinfo {author} {\bibfnamefont
  {T.}~\bibnamefont {Sasaki}},\ and\ \bibinfo {author} {\bibfnamefont
  {S.}~\bibnamefont {Iwai}},\ }\bibfield  {title} {\bibinfo {title}
  {{Collective Excitation of an Electric Dipole on a Molecular Dimer in an
  Organic Dimer-Mott Insulator}},\ }\href
  {https://doi.org/10.1103/PhysRevLett.110.106401} {\bibfield  {journal}
  {\bibinfo  {journal} {Phys. Rev. Lett.}\ }\textbf {\bibinfo {volume} {110}},\
  \bibinfo {pages} {106401} (\bibinfo {year} {2013})}\BibitemShut {NoStop}%
\bibitem [{\citenamefont {Kobayashi}\ \emph {et~al.}(2020)\citenamefont
  {Kobayashi}, \citenamefont {Ding}, \citenamefont {Taniguchi}, \citenamefont
  {Satoh}, \citenamefont {Kawamoto},\ and\ \citenamefont
  {Furukawa}}]{Kobayashi2020}%
  \BibitemOpen
  \bibfield  {author} {\bibinfo {author} {\bibfnamefont {T.}~\bibnamefont
  {Kobayashi}}, \bibinfo {author} {\bibfnamefont {Q.-P.}\ \bibnamefont {Ding}},
  \bibinfo {author} {\bibfnamefont {H.}~\bibnamefont {Taniguchi}}, \bibinfo
  {author} {\bibfnamefont {K.}~\bibnamefont {Satoh}}, \bibinfo {author}
  {\bibfnamefont {A.}~\bibnamefont {Kawamoto}},\ and\ \bibinfo {author}
  {\bibfnamefont {Y.}~\bibnamefont {Furukawa}},\ }\bibfield  {title} {\bibinfo
  {title} {{Charge disproportionation in the spin-liquid candidate
  $\ensuremath{\kappa}\ensuremath{-}{(\mathrm{ET})}_{2}{\mathrm{Cu}}_{2}{(\mathrm{CN})}_{3}$
  at 6 K revealed by $^{63}\mathrm{Cu}$ NQR measurements}},\ }\href
  {https://doi.org/10.1103/PhysRevResearch.2.042023} {\bibfield  {journal}
  {\bibinfo  {journal} {Phys. Rev. Research}\ }\textbf {\bibinfo {volume}
  {2}},\ \bibinfo {pages} {42023} (\bibinfo {year} {2020})}\BibitemShut
  {NoStop}%
\bibitem [{\citenamefont {Kawasugi}\ \emph {et~al.}(2019)\citenamefont
  {Kawasugi}, \citenamefont {Seki}, \citenamefont {Tajima}, \citenamefont {Pu},
  \citenamefont {Takenobu}, \citenamefont {Yunoki}, \citenamefont {Yamamoto},\
  and\ \citenamefont {Kato}}]{Kawasugi2019}%
  \BibitemOpen
  \bibfield  {author} {\bibinfo {author} {\bibfnamefont {Y.}~\bibnamefont
  {Kawasugi}}, \bibinfo {author} {\bibfnamefont {K.}~\bibnamefont {Seki}},
  \bibinfo {author} {\bibfnamefont {S.}~\bibnamefont {Tajima}}, \bibinfo
  {author} {\bibfnamefont {J.}~\bibnamefont {Pu}}, \bibinfo {author}
  {\bibfnamefont {T.}~\bibnamefont {Takenobu}}, \bibinfo {author}
  {\bibfnamefont {S.}~\bibnamefont {Yunoki}}, \bibinfo {author} {\bibfnamefont
  {H.~M.}\ \bibnamefont {Yamamoto}},\ and\ \bibinfo {author} {\bibfnamefont
  {R.}~\bibnamefont {Kato}},\ }\bibfield  {title} {\bibinfo {title}
  {{Two-dimensional ground-state mapping of a Mott-Hubbard system in a flexible
  field-effect device}},\ }\href {https://doi.org/10.1126/sciadv.aav7282}
  {\bibfield  {journal} {\bibinfo  {journal} {Sci. Adv.}\ }\textbf {\bibinfo
  {volume} {5}},\ \bibinfo {pages} {eaav7282} (\bibinfo {year}
  {2019})}\BibitemShut {NoStop}%
\bibitem [{\citenamefont {Kawasugi}\ and\ \citenamefont
  {Yamamoto}(2021)}]{Kawasugi2021}%
  \BibitemOpen
  \bibfield  {author} {\bibinfo {author} {\bibfnamefont {Y.}~\bibnamefont
  {Kawasugi}}\ and\ \bibinfo {author} {\bibfnamefont {H.~M.}\ \bibnamefont
  {Yamamoto}},\ }\bibfield  {title} {\bibinfo {title} {{Simultaneous control of
  Bandfilling and Bandwidth in the Electric-Double-Layer Transistor Based on
  the Organic Mott Insulator $\kappa$-(BEDT-TTF)$_{2}$Cu[N(CN)$_{2}$]Cl}},\
  }\href@noop {} {\bibfield  {journal} {\bibinfo  {journal} {Crystals}\ \textbf {\bibinfo {volume} {12}},\
  \bibinfo {pages} {42}} (\bibinfo {year} {2022})}\BibitemShut {NoStop}%
\bibitem [{\citenamefont {Jeschke}\ \emph {et~al.}(2012)\citenamefont
  {Jeschke}, \citenamefont {de~Souza}, \citenamefont {Valent{\'{i}}},
  \citenamefont {Manna}, \citenamefont {Lang},\ and\ \citenamefont
  {Schlueter}}]{Jeschke2012}%
  \BibitemOpen
  \bibfield  {author} {\bibinfo {author} {\bibfnamefont {H.~O.}\ \bibnamefont
  {Jeschke}}, \bibinfo {author} {\bibfnamefont {M.}~\bibnamefont {de~Souza}},
  \bibinfo {author} {\bibfnamefont {R.}~\bibnamefont {Valent{\'{i}}}}, \bibinfo
  {author} {\bibfnamefont {R.~S.}\ \bibnamefont {Manna}}, \bibinfo {author}
  {\bibfnamefont {M.}~\bibnamefont {Lang}},\ and\ \bibinfo {author}
  {\bibfnamefont {J.~A.}\ \bibnamefont {Schlueter}},\ }\bibfield  {title}
  {\bibinfo {title} {{Temperature dependence of structural and electronic
  properties of the spin-liquid candidate
  $\ensuremath{\kappa}$-(BEDT-TTF)${}_{2}$Cu${}_{2}$(CN)${}_{3}$}},\ }\href
  {https://link.aps.org/doi/10.1103/PhysRevB.85.035125} {\bibfield  {journal}
  {\bibinfo  {journal} {Phys. Rev. B}\ }\textbf {\bibinfo {volume} {85}},\
  \bibinfo {pages} {35125} (\bibinfo {year} {2012})}\BibitemShut {NoStop}%
\bibitem [{\citenamefont {Pustogow}\ \emph
  {et~al.}(2021{\natexlab{b}})\citenamefont {Pustogow}, \citenamefont
  {R{\"{o}}sslhuber}, \citenamefont {Tan}, \citenamefont {Uykur}, \citenamefont
  {B{\"{o}}hme}, \citenamefont {Wenzel}, \citenamefont {Saito}, \citenamefont
  {L{\"{o}}hle}, \citenamefont {H{\"{u}}bner}, \citenamefont {Kawamoto},
  \citenamefont {Schlueter}, \citenamefont {Dobrosavljevi´c},\ and\
  \citenamefont {Dressel}}]{Pustogow2021-percolation}%
  \BibitemOpen
  \bibfield  {author} {\bibinfo {author} {\bibfnamefont {A.}~\bibnamefont
  {Pustogow}}, \bibinfo {author} {\bibfnamefont {R.}~\bibnamefont
  {R{\"{o}}sslhuber}}, \bibinfo {author} {\bibfnamefont {Y.}~\bibnamefont
  {Tan}}, \bibinfo {author} {\bibfnamefont {E.}~\bibnamefont {Uykur}}, \bibinfo
  {author} {\bibfnamefont {A.}~\bibnamefont {B{\"{o}}hme}}, \bibinfo {author}
  {\bibfnamefont {M.}~\bibnamefont {Wenzel}}, \bibinfo {author} {\bibfnamefont
  {Y.}~\bibnamefont {Saito}}, \bibinfo {author} {\bibfnamefont
  {A.}~\bibnamefont {L{\"{o}}hle}}, \bibinfo {author} {\bibfnamefont
  {R.}~\bibnamefont {H{\"{u}}bner}}, \bibinfo {author} {\bibfnamefont
  {A.}~\bibnamefont {Kawamoto}}, \bibinfo {author} {\bibfnamefont {J.~A.}\
  \bibnamefont {Schlueter}}, \bibinfo {author} {\bibfnamefont {V.}~\bibnamefont
  {Dobrosavljevi´c}},\ and\ \bibinfo {author} {\bibfnamefont {M.}~\bibnamefont
  {Dressel}},\ }\bibfield  {title} {\bibinfo {title} {{Low-temperature
  dielectric anomaly arising from electronic phase separation at the Mott
  insulator-metal transition}},\ }\href
  {https://doi.org/10.1038/s41535-020-00307-0} {\bibfield  {journal} {\bibinfo
  {journal} {npj Quantum Mater.}\ }\textbf {\bibinfo {volume} {6}},\ \bibinfo
  {pages} {9} (\bibinfo {year} {2021}{\natexlab{b}})}\BibitemShut {NoStop}%
\end{thebibliography}
%

\acknowledgements
We acknowledge Y. Tan, V. Dobrosavljevi\'c, M. Dressel, R. Kato and H.M. Yamamoto for valuable discussions. This work was supported by JSPS KAKENHI, Grant Numbers JP16H06346, JP19K03730, JP19H00891.

\textbf{Author Contributions:}
Y.K. and H.S. performed the sample fabrication, cryogenic transport measurements and the data analyses. N.T. supervised the investigation. A.P. performed additional analyses, interpreted the results and conceptualized the manuscript. A.P. wrote the manuscript with inputs from Y.K. and N.T..
All authors discussed the experimental results. 

\textbf{Competing Interests:}
The authors declare no competing interests.


\end{document}